\documentclass[prx,nofootinbib,twocolumn,showkeys,superscriptaddress,preprintnumbers,floatfix]{revtex4-1}

\usepackage{dynlearn}

\usepackage{empheq}
\usepackage{perpage} \MakePerPage{footnote}

  \newcommand{\kB}{k_\text{B}}  

\begin{document}

\title{Thermodynamic Machine Learning through Maximum Work Production}

\author{Alexander B. Boyd}
\email{alecboy@gmail.com}
\affiliation{Complexity Institute, Nanyang Technological University, Singapore}
\affiliation{School of Physical and Mathematical Sciences, Nanyang Technological University, Singapore}

\author{James P. Crutchfield}
\email{chaos@ucdavis.edu}
\affiliation{Complexity Sciences Center and Physics Department,
University of California at Davis, One Shields Avenue, Davis, CA 95616}

\author{Mile Gu}
\email{mgu@quantumcomplexity.org}
\affiliation{Complexity Institute, Nanyang Technological University, Singapore}
\affiliation{School of Physical and Mathematical Sciences, Nanyang Technological University, Singapore}
\affiliation{Centre for Quantum Technologies, National University of Singapore, Singapore}

\date{\today}
\bibliographystyle{unsrt}

\begin{abstract}

Adaptive systems---such as a biological organism gaining survival advantage, an
autonomous robot executing a functional task, or a motor protein transporting
intracellular nutrients---must model the regularities and stochasticity in
their environments to take full advantage of thermodynamic resources.
Analogously, but in a purely computational realm, machine learning algorithms
estimate models to capture predictable structure and identify irrelevant noise
in training data. This happens through optimization of performance metrics,
such as model likelihood. If physically implemented, is there a sense in
which computational models estimated through machine learning are physically
preferred? We introduce the thermodynamic principle that work production is the
most relevant performance metric for an adaptive physical agent and compare
the results to the maximum-likelihood principle that guides machine learning.
Within the class of physical agents that most efficiently harvest energy from
their environment, we demonstrate that an efficient agent's model explicitly
determines its architecture and how much useful work it harvests from the
environment. We then show that selecting the maximum-work agent for given
environmental data corresponds to finding the maximum-likelihood model. This
establishes an equivalence between nonequilibrium thermodynamics and dynamic
learning. In this way, work maximization emerges as an organizing principle that
underlies learning in adaptive thermodynamic systems.

\end{abstract}


\preprint{arxiv.org:2006.15416}

\maketitle

\setstretch{1.1}
\section{Introduction}

What is the relationship, if any, between abiotic physical processes and
intelligence? Addressed to either living or artificial systems, this challenge
has been taken up by scientists and philosophers repeatedly over the last
centuries, from the 19$^{th}$ century teleologists \cite{Cuvi18a} and biological
structuralists \cite{Berg11a,Thom61} to cybernetics of the mid-20$^{th}$ century
\cite{Wien48,Wien88a} and contemporary neuroscience-inspired debates of the
emergence of artificial intelligence in digital simulations \cite{Denn91a}. The
challenge remains vital today \cite{Goul79a,Denn95a,Mayn98a,Wagn14a}. A key
thread in this colorful and turbulent history explores issues that lie
decidedly at the crossroads of thermodynamics and communication theory---of
physics and engineering. In particular, what bridges the dynamics of
the physical world and its immutable laws and principles to the purposeful
behavior intelligent agents? The following argues that an essential connector
lies in a new thermodynamic principle: work maximization drives learning.

Perhaps unintentionally, James Clerk Maxwell laid foundations for a physics of
intelligence with what Lord Kelvin (William Thomson) referred to as
``intelligent demons'' \cite{Thom74a}. Maxwell in his 1857 book \emph{Theory of
Heat} argued that a ``very observant'' and ``neat fingered being'' could
subvert the Second Law of Thermodynamics \cite{Maxw88a}. In effect, his
``finite being'' uses its intelligence (Maxwell's word) to sort fast from slow
molecules, creating a temperature difference that drives a heat engine to do
useful work. The demon presented an apparent paradox because directly
converting disorganized thermal energy to organized work energy is forbidden by
the Second Law. The cleverness in Maxwell's paradox turned on equating the
thermodynamic behavior of mechanical systems with the intelligence in an agent
that can accurately measure and control its environment. This established an
operational equivalence between energetic thermodynamic processes, on the one
hand, and intelligence, on the other.

\begin{figure}[tbp]
\centering
\includegraphics[width=\columnwidth]{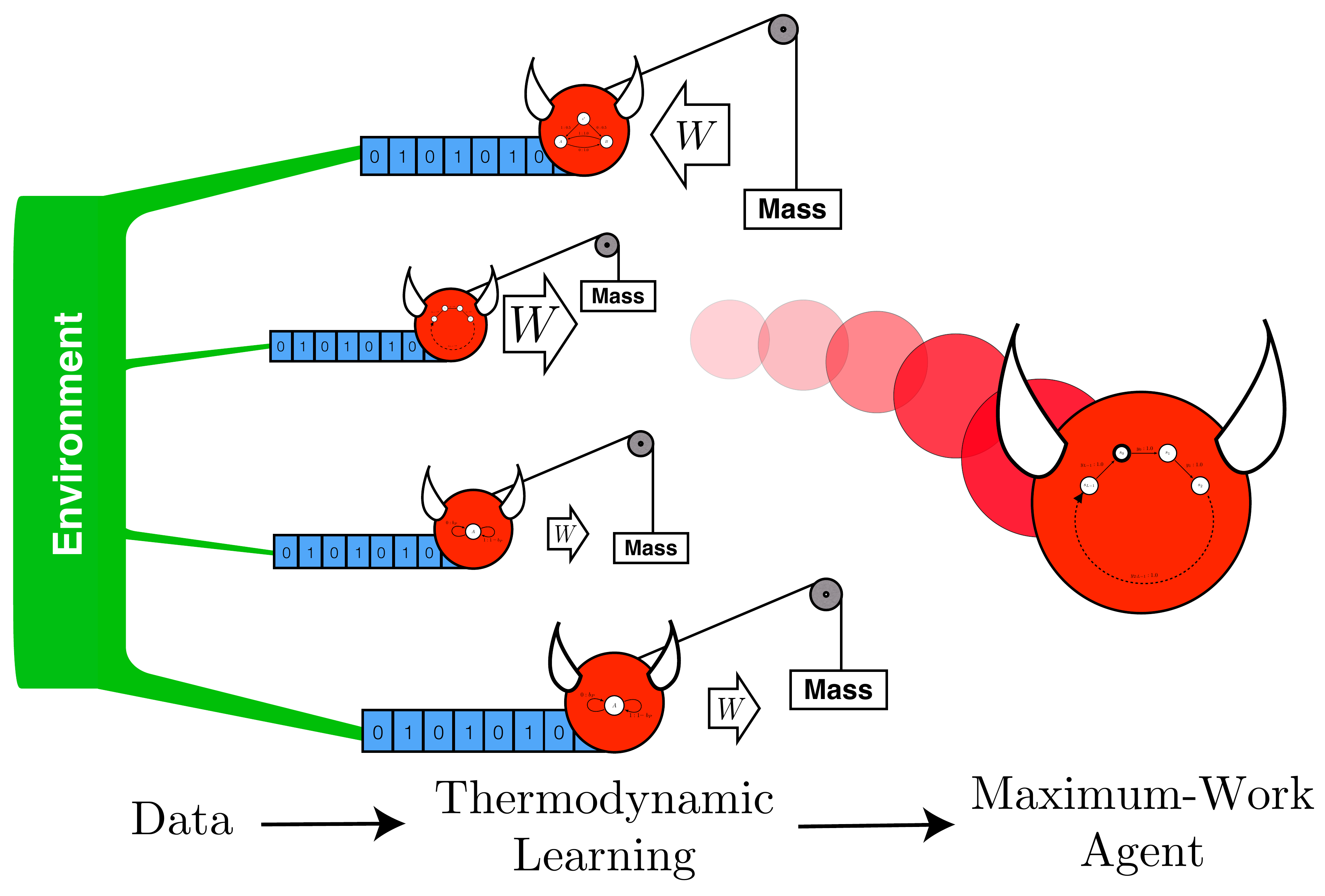}
\caption{Thermodynamic learning generates the maximum-work producing agent:
	(Left) Environment (green) behavior becomes data for agents (red). (Middle)
	Candidate agents each have an internal model (inscribed stochastic
	state-machine) that captures the environment's randomness and regularity to
	store work energy (e.g., lift a mass against gravity) or to borrow work
	energy (e.g., lower the mass). (Right) Thermodynamic learning searches the
	candidate population for the best agent---that producing the maximum work.
	}
\label{fig:ThermodynamicLearning}
\end{figure}

We will explore the intelligence of physical processes, substantially updating
the setting from the time of Kelvin and Maxwell, by calling on a wealth of
recent results on the nonequilibrium thermodynamics of information
\cite{Saga12a,Parr15a}. In this, we directly equate the operation of physical
agents descended from Maxwell's demon with notions of intelligence found in
modern machine learning. While learning is not necessarily the only capability
of a presumed intelligent being, it is certainly a most useful and interesting
feature.

The root of many tasks in machine learning lies in discovering structure from
data. The analogous process of creating models of the world from incomplete
information is essential to adaptive organisms, too, as they must model their
environment to categorize stimuli, predict threats, leverage opportunities, and
generally prosper in a complex world. Most prosaically, translating
\emph{training data} into a generative \emph{model} corresponds to \emph{density
estimation} \cite{Shai14a,Hast16a, Meht19a}, where the algorithm uses the data to
construct a probability distribution.

This type of model-building at first appears far afield from more familiar
machine learning tasks such as categorizing pet pictures into cats and dogs or
generating a novel image of a giraffe from a photo travelogue. Nonetheless, it
encompasses them both \cite{Lin17a}. Thus, by addressing thermodynamic roots of
model estimation, we seek a physical foundation for a wide breadth of machine
learning.

To carry out density estimation, machine learning invokes the principle of
maximum-likelihood to guide intelligent learning. This says, of the possible
models consistent with the training data, an algorithm should select that with
maximum probability of having generated the data. Our exploration of the
physics of learning asks whether a similar thermodynamic principle guides
physical systems to adapt to their environments.

The modern understanding of Maxwell's demon no longer entertains violating the
Second Law of Thermodynamics \cite{Land61a}. In point of fact, the Second Law's
primacy has been repeatedly affirmed in modern nonequilibrium theory and
experiment. That said, what has emerged is that we now understand how
intelligent (demon-like) physical processes can harvest thermal energy as
useful work. They do this by exploiting an information reservoir \cite{Benn87a,
Land61a, Szil29a}---a storehouse of information as randomness and correlation.
That reservoir is the demon's informational environment, and the mechanism by
which the demon measures and controls its environment embodies the demon's
intelligence, according to modern physics.  We will show that this mechanism is
directly linked to the demon's model of its environment, which allows us to
formalize the connection to machine learning.

Machine learning estimates different likelihoods of different models given the
same data.  Analogously, in the physical setting of information thermodynamics,
different demons harness different amounts of work from the same information
reservoir. Leveraging this commonality, Sec. \ref{sec: Framework} introduces
\emph{thermodynamic learning} as a physical process that infers optimal demons
from environmental information. As shown in Fig.
\ref{fig:ThermodynamicLearning}, thermodynamic learning selects demons that
produce maximum work, paralleling parametric density estimation's selection of
models with maximum likelihood. Section \ref{sec:Short Background} establishes
background in density estimation, computational mechanics, and thermodynamic
computing necessary to formalize the comparison of maximum-work and
maximum-likelihood learning. Our surprising result is that these two principles
of maximization are the same, when compared in a common setting.  This adds
credence to the longstanding perspective that thermodynamics and statistical
mechanics underlie many of the tools of machine learning \cite{Watk93a,
Enge01a, Bell11a, Dick15a, Gold17a, Alem18a, Meht19a, Bahr20a}.

Section \ref{sec:agent_energetics} formally establishes that to construct an
intelligent work-harvesting demon, a probabilistic model of its environment is
essential. That is, the demon's Hamiltonian evolution is directly determined by
its environmental model. This comes as a result of discarding demons that are
ineffective at harnessing energy from \emph{any} input, focusing only on a
refined class of \emph{efficient} demons that make the best use of the given
data. This leads to the central result, found in Sec. \ref{sec:Designing
Agents}, that the demon's work production from environmental ``training data''
increases linearly with the log-likelihood of the demon's model of its
environment. Thus, if the thermodynamic training process selects the
maximum-work demon for given data, it has also selected the maximum-likelihood
model for that same data.

Ultimately, our work demonstrates an \emph{equivalence between the conditions of
maximum work and maximum likelihood}. In this way, \emph{thermodynamic learning
is machine learning for thermodynamic machines}---it is a physical process that
infers models in the same way a machine learning algorithm does. Thus, work
itself can be interpreted as a \emph{thermodynamic performance measure for
learning}. In this framing, learning is physical, building on the long-lived
narrative of the thermodynamics of organization, which we recount in Sec.
\ref{sec:PrincOrg}. While it is natural to argue that learning confers
benefits, our result establishes that the benefit is fundamentally rooted in
the physical tradeoff between energy and information.

\section{Framework}
\label{sec: Framework}

While demons continue to haunt discussions of physical intelligence, the notion
of a physical process trafficking in information and energy exchanges need not
be limited to mysterious intelligent beings. Most prosaically, we are concerned
with any physical system that, while interacting with an environment,
simultaneously processes information at some energetic cost or benefit.
Avoiding theological distractions, we refer to these processes as
\emph{thermodynamic agents}. In truth, any physical system can be thought of as
an agent, but only a limited number of them are especially useful for or adept
at commandeering information to convert between various kinds of thermodynamic resources, such
as between heat and work. Here, we introduce a construction that
shows how to find physical systems that are the most capable of processing
information to affect thermodynamic transformations.

Consider an environment that produces information in the form of a time series
of physical values at regular time intervals of length $\tau$. We denote the
particular state realized by the environment's output at time $j\tau$ by the
symbol $y_j \in \mathcal{Y}_j$. Just as the agent must be instantiated by a
physical system, so too must the environment and its outputs to the agent.
Specifically, $\mathcal{Y}_j$ represents the state space of the $j$th output,
which is a subsystem of the environment.

An agent has no access to the internals of its environment and so treats it
as a black box. Thus, the agent can only access and interact with the
environment's output system $\mathcal{Y}_j$ over each time interval $t \in
(j\tau,(j+1)\tau)$. In other words, the state $y_j$ realized by the
environment's output is also the agent's \emph{input} at time $j\tau$.  For
instance, the environment may produce realizations of a two level spin system
$\mathcal{Y}_j=\{ \uparrow, \downarrow \}$, which the agent is then tasked to
manipulate through Hamiltonian control.

The aim, then, is to find an agent that produces as much work as possible using
these black-box outputs. To do so, the agent must come to know something about
the black box's structure. This is the \emph{principle of requisite complexity}
\cite{Boyd16d}---thermodynamic advantage requires that the agent's organization
match that of its environment. We implement this by introducing a method for
thermodynamic learning as shown in Fig. \ref{fig:ThermodynamicLearning}, that
selects a specific agent from a collection of candidates.

Peeking into the internal mechanism of the black box, we wait for a time
$L\tau$, receiving the $L$ symbols $y_{0:L}=y_0y_1 \cdots y_{L-1}$. This is the
agent's training data, which is copied as needed to allow a population of
candidate agents to interact with it. As each agent interacts with a copy, it
produces an amount of work, which it stores in the work reservoir for later use. 
In Fig. \ref{fig:ThermodynamicLearning}, the work reservoir is illustrated by a hanging 
mass which raises when positive work is produced, storing more energy in gravitational 
potential energy, and lowers when work production is negative, expending that same 
potential energy.  However the work energy is stored, after the agents harvest work from 
the training data, the agent that produced the most work is selected.

\emph{Finding the maximum-work agent is ``thermodynamic learning'' in the sense that it selects a device based
on measuring its thermodynamic performance---the amount of work the device
extracts.} Ultimately, the goal is that the agent selected by thermodynamic
learning continues to extract work as the environment produces new symbols.
However, we leave analyzing the long-term effectiveness of thermodynamic
learning to the future. Here, we concentrate on the condition of maximum-work
itself, deriving and interpreting it.

Section \ref{sec:agent_energetics} begins by describing the general class of
physical agents that can harness work from symbol sequence, known as
\emph{information ratchets} \cite{Mand012a, Boyd15a}. While these agents are
sufficiently general to implement virtually any (Turing) computation,
maximizing work production precludes a wide array of agents. Section
\ref{sec:Thermodynamically Efficient Computation} then refines our
consideration to agents that waste as little work as possible and, in so doing,
vastly narrow the search by thermodynamic learning. For this refined class of
agents, we find that each agent's operation is exactly determined by its
environment model. This leads to our final result, that the agent's work
increases linearly with the model's log-likelihood.

For clarity, note that thermodynamic learning differs from physical systems
that, evolving in time, dynamically adapt to their environment
\cite{Gold17a, Gold19a, Zhon20a}. Work maximization as described here is
thermodynamic in its objective, while these previous approaches to learning
are thermodynamic in their mechanism.

That said, the perspectives are linked. In particular, it was suggested that
physical systems spontaneously decrease work absorbed from driving
\cite{Gold19a}. Note that work absorbed by the system is opposite the work
produced. And so, as they evolve over time, these physical systems appear to
seek higher work production, paralleling how thermodynamic learning selects for
the highest work production. And, the synchronization by which a physical
system decreases work absorption is compared to learning \cite{Gold19a}.
Reference \cite{Zhon20a} goes further, comparing the effectiveness of physical
evolution to maximum-likelihood estimation employing an autoencoder. Notably,
it reports that that form of machine learning performs markedly better than
physical evolution, for the particular system considered there. By contrast, we
show that the advantage of machine learning over thermodynamic learning does
not hold in our framework. Simply speaking, they are synonymous.

We compare thermodynamic learning to machine learning algorithms that use
maximum-likelihood to select models consistent with given data. As Fig.
\ref{fig:ThermodynamicLearning} indicates, each agent has an internal model of
its environment; a connection Sec. \ref{Agent--Model Equivalence} formalizes.
Each agent's work production is then evaluated for the training data. Thus,
arriving at a maximum-work agent also selects that agent's internal model as a
description of the environment. Moreover and in contrast with Ref.
\cite{Zhon20a}, which compares thermodynamic and machine learning methods
numerically, the framework here leads to an analytic derivation of the
equivalence between thermodynamic learning and maximum-likelihood density
estimation.

\section{Preliminaries}
\label{sec:Short Background}

Directly comparing thermodynamic learning and density estimation requires
explicitly demonstrating that thermodynamically-embedded computing and machine
learning share the framework just laid out. The following introduces what we
need for this: concepts from machine learning, computational mechanics, and
thermodynamic computing. (Readers preferring fuller detail should refer to App.
\ref{app:Extended Background}.)

\subsection{Parametric Density Estimation}
\label{sec:DensityEst}

Parametric estimation determines, from training data, the parameters $\theta$
of a probability distribution. In the present setting, $\theta$ parametrizes a
family of probabilities $\Pr(Y_{0:\infty}=y_{0:\infty}|\Theta=\theta)$ over
sequences (or \emph{words}) of any length. Here, $Y_{0:\infty}=Y_0Y_1 \cdots $
is the infinite-sequence random variable, composed of the random variables
$Y_j$ that each realize the environment's output $y_j$ at time time $j\tau$,
and $\Theta$ is the random variable for the model.  In other words, a given
model $\theta$ predicts the probability of any sequence $y_{0:L}$ of any length
$L$ that one might see.

For convenience, we introduce random variables $Y^\theta_{j}$ that define a
model:
\begin{align*}
\Pr(Y^\theta_{0:\infty})\equiv \Pr(Y_{0:\infty}|\Theta=\theta)
  ~.
\end{align*}
With training data $y_{0:L}$, the \emph{likelihood} of model $\theta$ is given
by the probability of the data given the model:
\begin{align*}
\mathcal{L}(\theta|y_{0:L}) & = \Pr(Y_{0:L}=y_{0:L}|\Theta=\theta) \\
  & = \Pr(Y^\theta_{0:L}=y_{0:L})
  ~.
\end{align*}
Parametric density estimation seeks to optimize the likelihood
$\mathcal{L}(\theta|y_{0:L})$ \cite{Shai14a, Reze14a, Meht19a}. However, the
procedure that finds maximum-likelihood estimates usually employs the
\emph{log-likelihood} instead:
\begin{align}
\ell (\theta|y_{0:L})=\ln \Pr(Y^\theta_{0:L}=y_{0:L})
  ~,
\label{eq:defLogLikelihood}
\end{align}
since it is maximized for the same models, but converges more effectively
\cite{Demp77a}.

\subsection{Computational Mechanics}
\label{sec:CMech}

Given that our data is a time series of arbitrary length starting with $y_0$, we
must choose a model class whose possible parameters $\mathbf{\Theta} =
\{\theta\}$ specify a wide range of possible distributions
$\Pr(Y^\theta_{0:\infty})$---the \emph{semi-infinite processes}. \EMs, a class
of finite-state machines introduced to describe bi-infinite processes
$\Pr(Y^\theta_{-\infty:\infty})$, provide a systematic means to do this
\cite{Crut12a}. As described in App. \ref{app:Extended Background} these
finite-state machines comprise just such a flexible class of representations;
they can describe any semi-infinite process. This follows from the fact that
they are the minimal sufficient statistic for prediction explicitly constructed
from the process.

\begin{figure}[tbp]
\centering
\includegraphics[width=.6\columnwidth]{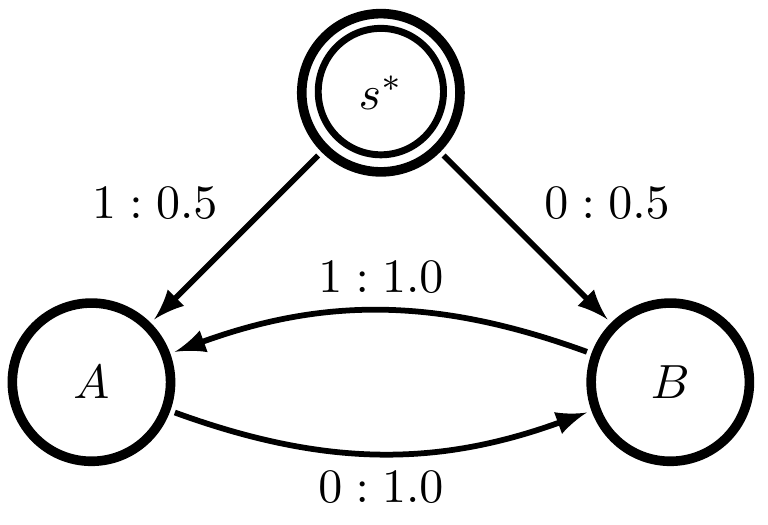}
\caption{\EM\ generating the phase-uncertain period-$2$ process: With
	probability $0.5$, an initial transition is made from the start state $s^*$
	to state $A$. From there, it emits the sequence $1010 \ldots$. However, with
	probability $0.5$, the start state transitions to state $B$ and outputs
	the sequence $0101 \ldots$.
	}
\label{fig:HMM}
\end{figure}

A process's \eM\ consists of a set of hidden states $\mathcal{S}$, a set of
output states $\mathcal{Y}$, a start state $s^* \in \mathcal{S}$, and
conditional output-labeled transition matrix $\theta^{(y)}_{s \rightarrow s'}$
over the hidden states:
\begin{align*}
\theta^{(y)}_{s \rightarrow s'}
  = \Pr(S^\theta_{j+1}=s',Y^\theta_j=y|S^\theta_j=s)
  ~.
\end{align*}
$\theta^{(y)}_{s \rightarrow s'}$ specifies the probability of transitioning to
hidden state $s'$ and emitting symbol $y$ given that the machine is in state
$s$. In other words, the model is fully specified by the tuple:
\begin{align*}
\theta = \{\mathcal{S},\mathcal{Y},s^*,\{\theta^{(y)}_{s \rightarrow s'}\}_{s,s' \in \mathcal{S},y \in \mathcal{Y}}\}
  ~.
\end{align*}
As an example, Fig. \ref{fig:HMM} shows an \eM\ that generates a periodic
process with initially uncertain phase.

\EMs\ are \emph{unifilar}, meaning that the current causal state $s_j$ along
with the next $k$ symbols uniquely determines the following causal state
through the \emph{propagator function}:
\begin{align*}
s_{j+k}=\epsilon(s_j,y_{j:j+k})
  ~.
\end{align*}
This yields a simple expression for the probability of any word in terms of the
model parameters:
\begin{align*}
\Pr(Y^\theta_{0:L}=y_{0:L})
  = \prod_{j=0}^{L-1}
  \theta^{(y_j)}_{\epsilon(s^*,y_{0:j})\rightarrow \epsilon(s^*,y_{0:j+1})}
  ~.
\end{align*}
In addition to being uniquely determined by the semi-infinite process,
the \eM\ uniquely generates that same process. This means that our model class
$\Theta$ is equivalent to the class of possible distributions over time series
data.  Moreover, knowledge of the causal state of an \eM\ at any time step $j$
contains all information about the future that could be predicted from the
past. In this sense, the causal state is \emph{predictive} of the process.
These and other properties have motivated a long investigation of \eMs,
in which the memory cost of storing the causal states is frequently
used as a measure of process structure. Appendix \ref{app:Extended Background} gives an extended review.

\subsection{Thermodynamic Computing}
\label{subsec:Thermodynamic Computing}

Computation is physical---any computation takes place embedded in a physical
system. Here, we refer to substrate of the physically-embedded computation as
the \emph{system of interest} (SOI). Its states, denoted $\mathcal{Z}=\{z\}$,
are taken as the underlying physical system's \emph{information bearing degrees
of freedom} \cite{Land61a}. The SOI's dynamic evolves the state distribution
$\Pr(Z_t=z_t)$, where $Z_t$ is the random variable describing state at time
$t$. Computation over time interval $t \in [\tau,\tau']$ specifies how the
dynamic maps the SOI from the initial time $t=\tau$ to the final time
$t=\tau'$. It consists of two components:
\begin{enumerate}
      \setlength{\topsep}{0pt}
      \setlength{\itemsep}{0pt}
      \setlength{\parsep}{0pt}
\item An initial distribution over states $\Pr(Z_\tau=z_\tau)$ at time $t=\tau$.
\item Application of a Markov channel $M$, characterized by the conditional
	probability of transitioning to the final state $z_{\tau'}$ given the
	initial state $z_\tau$:
\begin{align*}
M_{z_\tau \rightarrow z_{\tau'}}= \Pr(Z_{\tau'}=z_{\tau'}|Z_\tau=z_\tau)
  ~.
\end{align*}
\end{enumerate}
Together, these specify the SOI's computational elements. In this, $z_\tau$ is
the \emph{input} to the physical computation, $z_{\tau'}$ is the \emph{output},
and $M_{z_\tau \rightarrow z_\tau'}$ is the \emph{logical architecture}.

\begin{figure}[tbp]
\centering
\includegraphics[width=\columnwidth]{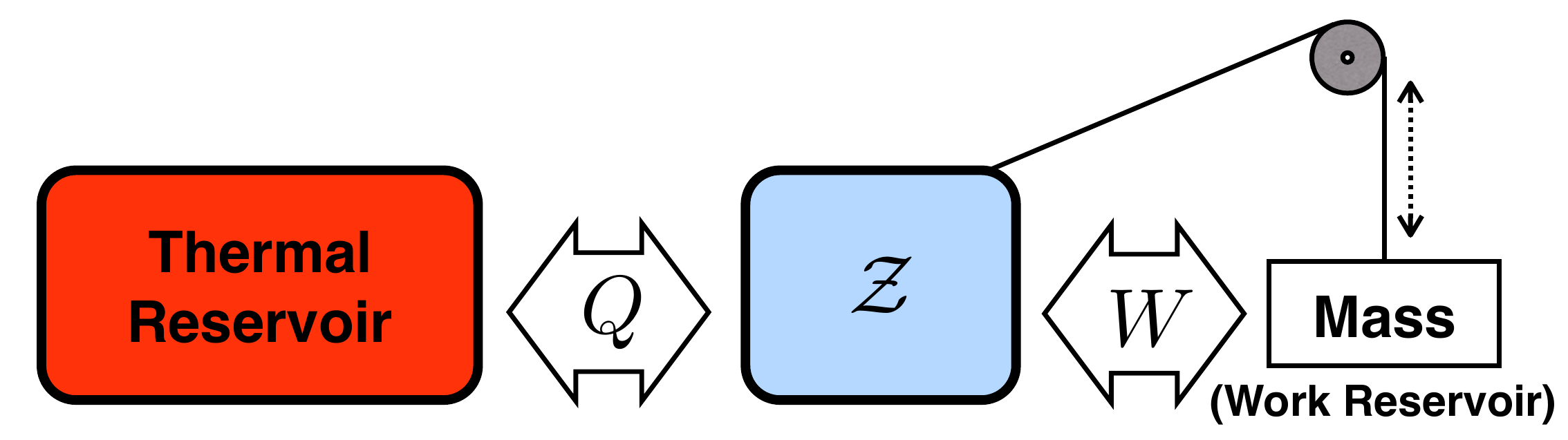}
\caption{Thermodynamic computing: The system of interest $\mathcal{Z}$'s states
	store information, processing it as they evolve. The work reservoir,
	represented as the suspended mass, supplies work energy $W$ to drive the
	SOI Hamiltonian along a deterministic trajectory
	$\mathcal{H}_\mathcal{Z}(t)$. Meanwhile, heat energy $Q$ is exchanged with
	the thermal reservoir, driving the system toward thermal equilibrium.
	}
\label{fig:PhysicalComputation}
\end{figure}

Figure \ref{fig:PhysicalComputation} illustrates a computation's physical
implementation. SOI $\mathcal{Z}$ is coupled to a work reservoir, depicted as a
mass hanging from a string, that controls the system's Hamiltonian along a
trajectory $\mathcal{H}_\mathcal{Z}(t)$ over the computation interval $t \in
[\tau,\tau']$ \cite{Deff2013}. This is the basic definition of a
\emph{thermodynamic agent}: an evolving Hamiltonian driving a physical system
to compute at the cost of work.

In a classical system, this control determines each state's energy $E(z,t)$. As
a result of the control, changes in energy due to changes in the Hamiltonian
correspond to work exchanges between the SOI and work reservoir. The system
$\mathcal{Z}$ follows a \emph{state trajectory} $z_{\tau:\tau'}$ over the time
interval $t \in [\tau, \tau']$, which we can write:
\begin{align*}
z_{\tau:\tau'} = z_\tau z_{\tau+dt} \cdots z_{\tau'-dt} z_{\tau'}
  ~,
\end{align*}
where $z_t$ is the system state at time $t$. Here, we decomposed the trajectory
into intervals of duration $dt$, taken short enough to yield infinitesimal
changes in state probabilities and the Hamiltonian. The resulting \emph{work
production} for this trajectory is then the integrated change in energy due to
the Hamiltonian's time dependence \cite{Deff2013}:
\begin{align*}
 W_{|z_{\tau:\tau'}}  =- \int_{\tau}^{\tau'} dt\, \partial_t
 E(z,t)\big\vert_{z=z_t}
  ~.
\end{align*}

Note that while the state trajectory $z_{\tau:\tau'}$ mirrors the time series
notation used for the training data $y_{0:L}=y_0y_1 \cdots y_{L-1}$, they are
different objects and should not be conflated. On the one hand, the training
data series $y_{0:L}$ is composed of realizations of $L$ \emph{separate}
subsystems, each produced at different times $j\tau$, $j \in \{0,1,2,\cdots L-1
\}$.  $y_j$ is realized in the subsystem $\mathcal{Y}_j$, and so it can be
manipulated completely separately from any other element of $y_{0:L}$ lying
outside of $\mathcal{Y}_j$. By contrast, $z_t$ depends dynamically on many
other elements in $z_{\tau:\tau'}$, all of which lie in the \emph{same} system,
since the time series $z_{\tau:\tau'}$ represents state evolution of the
single system $\mathcal{Z}$ over time.

While the SOI exchanges work energy with the work reservoir, as Fig.
\ref{fig:PhysicalComputation} shows, it exchanges heat $Q$ with the thermal
reservoir. Coupling to a heat reservoir adds stochasticity to the state
trajectory $z_{\tau:\tau'}$.
Since the SOI computes while coupled to a thermal reservoir at temperature $T$,
\emph{Landauer's Principle} \cite{Land61a} relates a computation's logical
processing to its energetics. In its contemporary form, it bounds the
\emph{average} work production $\langle W \rangle$ by a term proportional to
SOI's entropy change.  Taking the Shannon entropy $H[Z_t]=-\sum_{z} \Pr(Z_t=z)
\ln \Pr(Z_t=z)$ in natural units, the Second Law of Thermodynamics implies
\cite{Parr15a}:
\begin{align*}
\langle W \rangle \leq \kB T \left(H[Z_{\tau'}]-H[Z_{\tau}] \right)
  ~.
\end{align*}
Here, the average $\langle W \rangle$ is taken over all possible microscopic
trajectories. And, the energy landscape is assumed to be flat at the
computation's start and end, giving no energetic preference to a particular
informational state.

\section{Agent Energetics}
\label{sec:agent_energetics}

We now construct the theoretical framework for how agents extract work from
time-series data. This involves breaking down the agent's actions into
manageable elementary components---where we demonstrate their actions can be
described as repeated application of sequence of computations. We then introduce
tools to analyze work production within such general computations on finite
data. We highlight the importance of the agent's model of the data in
determining work production. This model-dependence emerges by refining the
class of agents to those that execute their computation most efficiently. The
results are finally combined, resulting in a closed-form expression for agent
work production from time-series data.

\subsection{Agent Architecture}
\label{sec:Work Production of Thermodynamic Agents}

Recall from Sec. \ref{sec: Framework} that the basic framework describes
a thermodynamic agent interacting with an environment at regular time-intervals
$\tau j$ in state $y_j$. Each $y_j$ is drawn according to a random variable
$Y_j$, such that the sequence $Y_{0:\infty} = Y_0Y_1\ldots$ is a
semi-infinite stochastic process. The agent's task is to interact with
this input string to generate useful work.

For example, consider an agent charged with extracting work from an alternating
process---a sequence emitted by a degenerate two-level system that alternates
periodically between symbols $0$ and $1$. In isolation each symbol looks
random and has no free energy. Thus, an agent that interacts with each symbol
the same way gains no work. However, a memoryful agent can adaptively adjust
its behavior, after reading the first symbol, to exactly predict succeeding
symbols and, therefore, extract meaningful work. This method of harnessing
\emph{temporal correlations} is implemented by \emph{information ratchets}
\cite{Mand012a,Boyd15a}. They combine physical inputs with additional agent
memory states that store the input's temporal correlations.

\begin{figure}[tbp]
\centering
\includegraphics[width=\columnwidth]{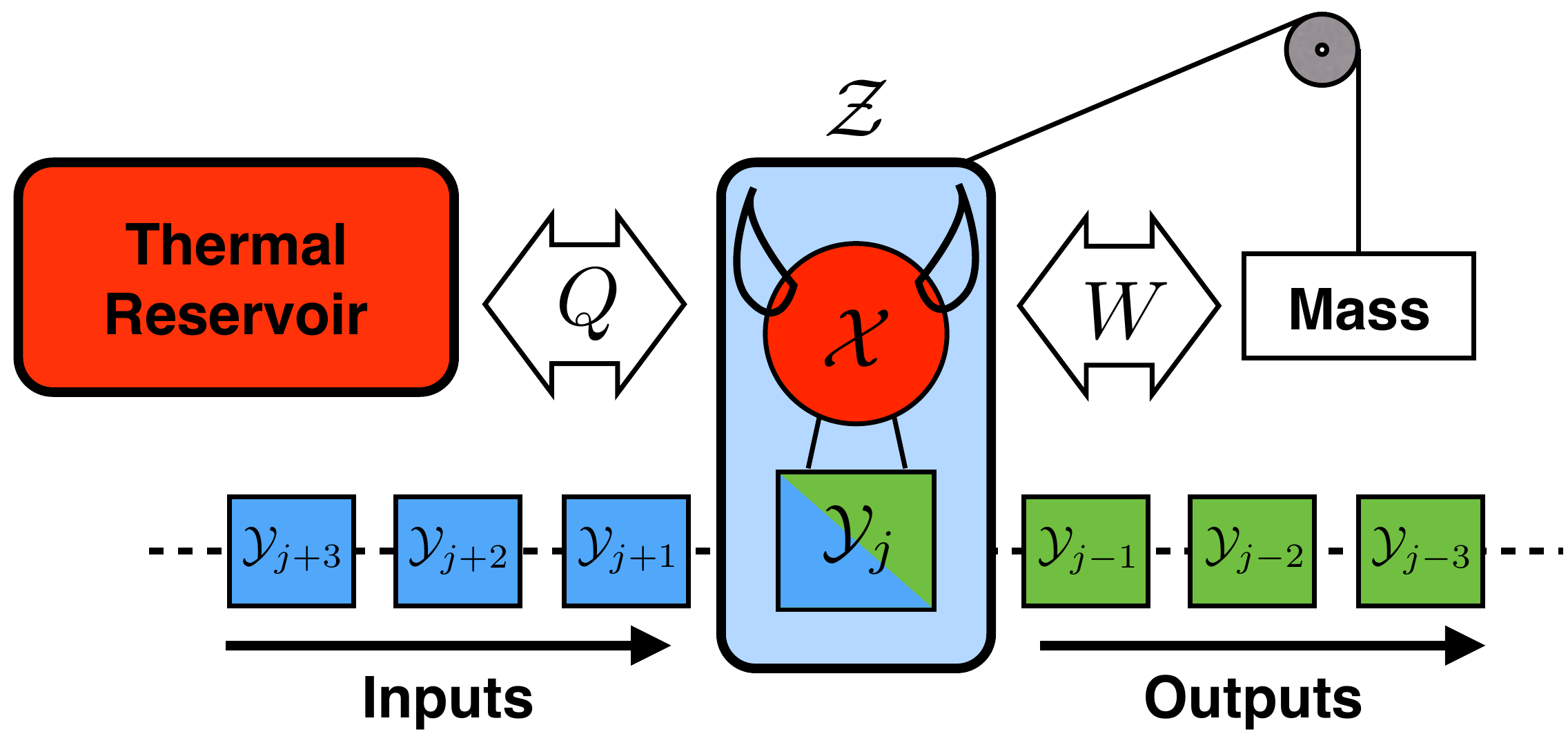}
\caption{Thermodynamic computing by an agent driven by an input sequence:
	Information bearing degrees of freedom of SOI $\mathcal{Z}$ in the $j$th
	interaction interval split into the direct product of agent states
	$\mathcal{X}$ and the $j$th input states $\mathcal{Y}_j$. Work $W$ and heat
	$Q$ are defined in the same way as in Fig. \ref{fig:PhysicalComputation},
	with the SOI's Hamiltonian control $\mathcal{H}_{\mathcal{X} \times
	\mathcal{Y}_j}(t)$ explicitly decoupled from the environment's remaining
	subsystems $\cdots
	\mathcal{Y}_{j-2}\mathcal{Y}_{j-1}\mathcal{Y}_{j+1}\mathcal{Y}_{j+2}
	\cdots$, corresponding to future inputs $\mathcal{Y}_{j+1}\mathcal{Y}_{j+2}
	\cdots$ and past outputs $ \cdots \mathcal{Y}_{j-2}\mathcal{Y}_{j-1}$.
	}
\label{fig:AgentAndBit}
\end{figure}

As shown in Fig. \ref{fig:AgentAndBit}, we describe an agent's memory via an ancillary physical system $\mathcal{X}$.  The agent then operates cyclically with duration $\tau$, such that the $j^{th}$
cycle runs over the time-interval $[j \tau, (j + 1) \tau )$. Each cycle
involves two phases:
\begin{enumerate}
      \setlength{\topsep}{-2pt}
      \setlength{\itemsep}{-4pt}
      \setlength{\parsep}{-2pt}
\item \emph{Interaction}: Agent memory $\mathcal{X}$ couples to and interacts
	with the $j^{th}$ input system $\mathcal{Y}_j$ that contains the $j^{th}$
	input symbol $y_j$. This phase has duration $\tau' < \tau$, meaning the
	$j$th interaction phase occurs over the time-interval $[j \tau, j
	\tau+\tau')$. At the end, the agent decouples from the system
	$\mathcal{Y}_j$, passing its new state $y'_j$ to the environment as output or
	exhaust.
\item \emph{Rest}: During time interval $[j\tau+\tau',(j+1)\tau)$, the
	agent's memory $\mathcal{X}$ sits idle, waiting for the next input
	$\mathcal{Y}_{j+1}$.
\end{enumerate}
In this way, the agent transforms a series of inputs $y_{0:L}$ into a series of
outputs $y_{0:L}'$. 

In each cycle, all nontrivial thermodynamics occur in the interaction phase,
during which the SOI consists of the joint agent-input system: i.e.,
$\mathcal{Z} = \mathcal{X} \otimes \mathcal{Y}_j$, as shown in Fig.
\ref{fig:AgentAndBit}.  While the other subsystems $\cdots \mathcal{Y}_{j-2}
\mathcal{Y}_{j-1}$ and $\mathcal{Y}_{j+1}\mathcal{Y}_{j+2} \cdots$ may be
physically instantiated somewhere else in the environment, they do not
participate in the interaction, since they are energetically decoupled from
the agent in this phase. Paralleling the computation shown in Fig.
\ref{fig:PhysicalComputation}, Hamiltonian control over the joint space
$\mathcal{H}_{\mathcal{X}\times\mathcal{Y}_j}(t)$ results in a transformation
of the agent's SOI that also requires the exchange of work and heat.

\begin{figure}[tbp]
\includegraphics[width=\columnwidth]{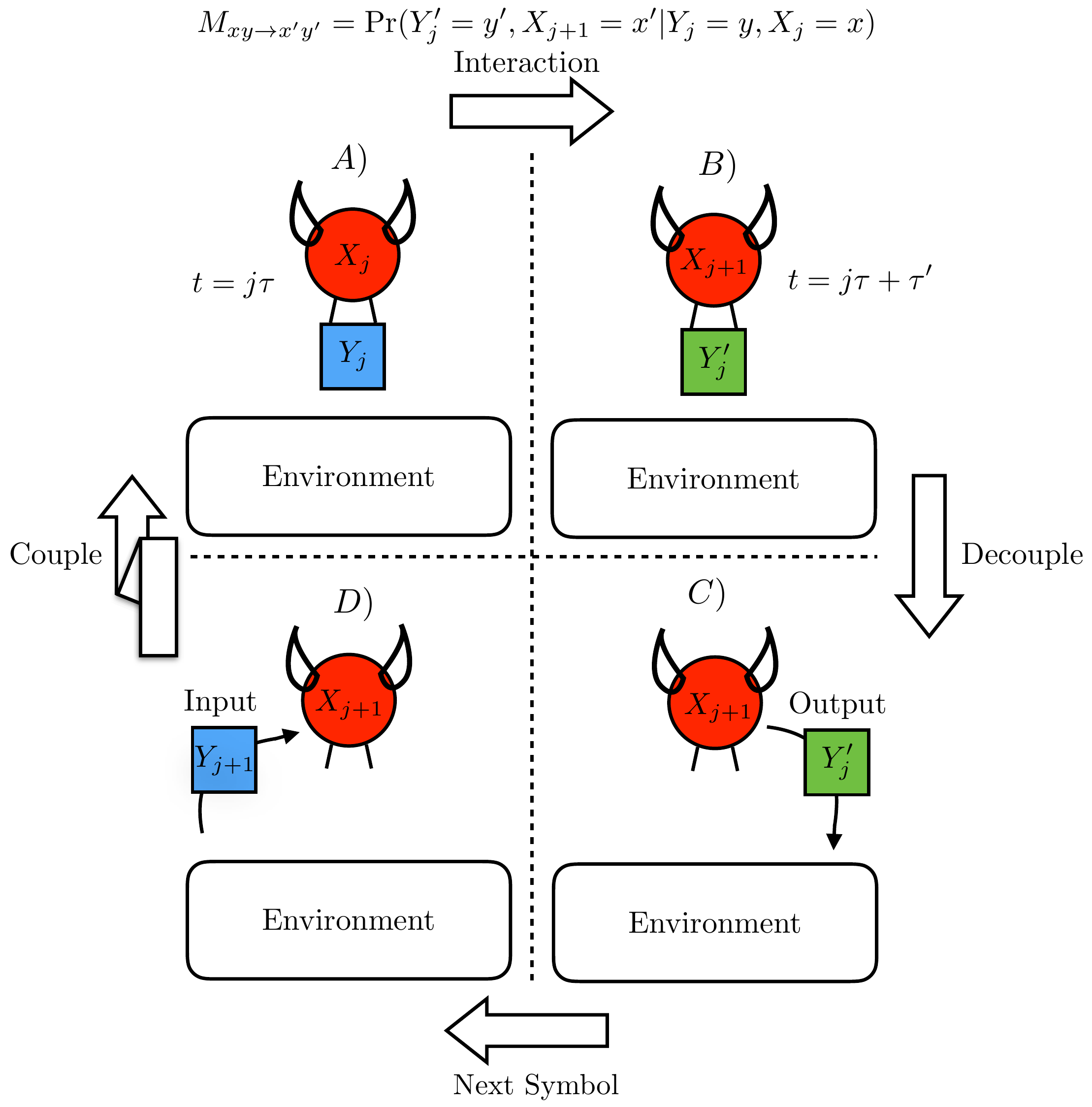}
\caption{Agent interacting with an environment via repeated symbol exchanges:
	A) At time $j\tau$ agent memory $X_j$ begins interacting with input
	symbol $Y_j$. Transitioning from A) to B), agent memory and
	interaction symbol jointly evolve according to the Markov channel
	$M_{xy\rightarrow x'y'}$. This results in B)---the updated states of agent
	memory $X_{j+1}$ and interaction symbol $Y_{j}'$ at time $j\tau+\tau'$.
	Transitioning from B) to C), the agent memory decouples from the
	interaction symbol, emitting its new state to the environment. Then,
	transitioning from C) to D), the agent retains its memory state
	$X_{j+1}$ and the environment emits the next interaction symbol $Y_{j+1}$.
	Finally, transitioning from D) to A), the agent restarts the cycle by
	coupling to the next input symbol.
	}
\label{fig:ThermodynamicAgent}
\end{figure}

This interaction phase updates SOI states according to a Markov transition
matrix $M$ shown in Fig. \ref{fig:ThermodynamicAgent}:
\begin{align}
M_{xy \rightarrow x' y'}
  \! = \! \Pr(X_{j+1} \! = \! x',Y'_j=y'|X_j \! = \!  x,Y_j \! = \! y)
  ~,
\label{eq:JointMarkov}
\end{align}
where $X_j$ and $X_{j+1}$ are the random variables for the states of the
agent's memory $\mathcal{X}$ before and after the $j$th interaction interval,
and $Y_j$ and $Y'_j$ are the random variables for the system $\mathcal{Y}_j$
before and after the same interaction interval, realizing the input and output,
respectively.

As Sec. \ref{subsec:Thermodynamic Computing} described, $M$ is the logical
architecture of the physical computation that transforms the agent's memory and
input simultaneously.  It is the central element in the agent's procedure for
transforming inputs $y_{0:L}$ into associated outputs $y'_{0:L}$. The key
observation is that $M$ captures all of the agent's internal logic. The logic
does not change from cycle to cycle. However, the presence of persistent
internal memory between cycles implies that the agent's behavior adapts to past
inputs and outputs. This motivates us to define $M$ as the \emph{agent
architecture} since it determines how an agent stores information temporally.
As we will show, $M$ is one of two essential elements in determining the work
an agent produces from a time series.

Note that prior related efforts to address agent energetics focused on
ensemble-average work production
\cite{Mand012a,Boyd15a,Boyd16c,Boyd16d,Boyd16e,Merh15a,Merh17a,Garn15}. In
contrast, here we relate work production to parametric density
estimation---which involves each agent being given a specific data string
$y_{0:L}$ for training. To address this case, the following determines the work
production for single-shot short input strings.

\subsection{Energetics of Computational Maps}
\label{sec:Thermodynamically Efficient Computation}

The agent architecture $M$ specifies a physical computation as described in Sec.
\ref{subsec:Thermodynamic Computing} and therefore has a minimum energy cost
determined by Landauer's bound. However, this is a bound on the \emph{average}
work production, which depends explicitly on the distribution of inputs. We
need to determine, instead, the work produced from a \emph{single} input $y_j$.
To find this we return to the general case of SOI $\mathcal{Z}$ undergoing a
thermodynamic computation $M$.

A physical operation takes the SOI from state $z_\tau$ at time $\tau$ to state
$z_{\tau'}$ at time $\tau'$. This specifies a \emph{computational map} $z_\tau
\rightarrow z_{\tau'}$ that ignores intermediate states in the SOI state
trajectory, as all information relevant to the computation's logical operation
lies in the input and output. Thus, our attention turns to the question: What
is the work production of a computational map $z_\tau \rightarrow z_{\tau'}$
performed by the computation $M$ at temperature $T$?

To determine this, we first prove a useful relation between the entropy and
work production for a particular state trajectory $z_{\tau:\tau'}$.
Specifically, let $W_{|z_{\tau:\tau'}}$ and $\Sigma_{|z_{\tau:\tau'}}$ denote
the work and total entropy production along this trajectory, respectively.
Meanwhile, let $E(z_t,t)$ denote the system energy when it is in state $z_t$ at
time $t$. Now, consider the \emph{pointwise nonequilibrium free energy}:
\begin{align}
\label{eq:pointfree}
\phi(z_t,t)=E(z_t,t)+ \kB T \ln \Pr(Z_t=z_t)
  ~.
\end{align}
More familiarly, note that its time-averaged quantity is the \emph{nonequilibrium free energy}
\cite{Deff12a}:
\begin{align*}
F^\text{neq}=\langle \phi(z,t)
\rangle_{\Pr(Z_t=z)}
  ~.
\end{align*}

We can then show that the \emph{entropy production} $\Sigma$ can be expressed:
\begin{align}
\Sigma_{|z_{\tau:\tau'}}
  & = \frac{-W_{|z_{\tau:\tau'}}+\phi(z_\tau,\tau)-\phi(z_{\tau'},\tau')}{T}
  ~.
\label{eq:relation1}
\end{align}
This follows by noting that the total entropy produced from thermodynamic
control is the sum of the entropy change in the system \cite{Espo11}:
\begin{align*}
\Delta S^\mathcal{Z}_{|z_{\tau:\tau'}}
  &= \kB \ln \frac{\Pr(Z_\tau=z_\tau)}{\Pr(Z_{\tau'}=z_{\tau'})}
\end{align*}
and that of the thermal reservoir:
\begin{align*}
\Delta S^\text{reservoir}_{|z_{\tau:\tau'}}&=\frac{Q_{|z_{\tau:\tau'}}}{T}
  ~.
\end{align*}
Equation (\ref{eq:relation1}) follows by summing up these contributions to the
total entropy production $\Sigma =\Delta S^\text{reservoir}+\Delta
S^\mathcal{Z}$ and noting that the SOI's change in energy obeys the First Law
of Thermodynamics $\Delta E^\mathcal{Z} = - W - Q$.

Since only the SOI's initial and final states matter to the logical operation
of the computational map, we take a statistical average of all trajectories
beginning in $z_\tau$ and ending in $z_{\tau'}$. This results in the work
production:
\begin{align}
\left\langle  W_{|z_\tau,z_{\tau'}} \right\rangle
   = \sum_{z'_{\tau:\tau'}}
  W_{|z'_{\tau:\tau'}}
  \Pr(Z_{\tau:\tau'}=z'_{\tau:\tau'}|z_\tau,z_{\tau'})
  ~,
\label{eq:ComputationalTrajectory}
\end{align}
for the computational map $z_\tau \rightarrow z_{\tau'}$. This determines how
much energy is stored in the work reservoir on average when a computation
results in this particular input-output pair.

Similarly, taking the same average of the entropy production shown in Eq. $(\ref{eq:relation1})$, conditioned on inputs and outputs, gives:
\begin{align*}
T \left\langle \Sigma_{|z_\tau,z_{\tau'}} \right\rangle
  & = -\langle W_{|z_\tau,z_{\tau'}} \rangle
  +\phi(z_\tau,\tau)-\phi(z_{\tau'},\tau'), \\
  & =-\langle W_{|z_\tau,z_{\tau'}} \rangle
  - \Delta \phi_{|z_\tau, z_{\tau'}}
  ~.
\end{align*}
This suggestively relates computational-mapping work and the change in
pointwise nonequilibrium free energy $\phi(z,t)$.

This relation between work and free energy simplifies for thermodynamically-efficient computations. In
such scenarios, the average total entropy production over all trajectories 
vanishes. Appendix \ref{app:zero_ent} shows that zero average entropy production, combined with the Crooks
fluctuation theorem~\cite{Croo99a, Jarz00a}, implies that entropy production along any
individual trajectory $z_{\tau:\tau'}$ produces zero entropy: $\Sigma_{|z_{\tau:\tau'}}=0$. This is expected from linear response \cite{Spec04a}.

Thus, substituting zero entropy production into Eq. (\ref{eq:relation1}), we
arrive at our result: \emph{work production for thermodynamically-efficient
computations is the change in pointwise nonequilibrium free energy}:
\begin{align*}
W^\text{eff}_{|z_{\tau:\tau'}}
  & =-\Delta \phi_{|z_\tau, z_{\tau'}}.
\end{align*}
Substituting Eq. (\ref{eq:pointfree}) then gives:
\begin{align*}
W^\text{eff}_{|z_{\tau:\tau'}}
  & = -\Delta E_\mathcal{Z}
  + \kB T \ln \frac{\Pr(Z_{\tau}=z_{\tau})}{\Pr(Z_{\tau'}=z_{\tau'})}
  ~,
\end{align*}
where $\Delta E_\mathcal{Z}=E(z_{\tau'},\tau')-E(z_{\tau},\tau)$. This is what
we would expect in quasistatic computations, where the system energies $E(z,t)$
are varied slowly enough that the system $\mathcal{Z}$ remains in equilibrium
for the duration. We should note, though, that it is possible to implement
efficient computations rapidly and out of equilibrium \cite{Ray20b}.

This also holds if we average over intermediate states of the SOI's state
trajectory, yielding the work production of a computational map:
\begin{align}
\left\langle W^\text{eff}_{|z_\tau,z_{\tau'}} \right\rangle
  = -\Delta E_\mathcal{Z}
  + \kB T \ln \frac{\Pr(Z_{\tau}=z_{\tau})}{\Pr(Z_{\tau'}=z_{\tau'})}
  ~.
\label{eq:EfficientWork2}
\end{align}
The energy required to perform efficient computing is independent of
intermediate properties. It depends only on the probability and energy of
initial and final states. This measures the energetic gains from a single data
realization as it transforms during a computation, as opposed to the ensemble
average.

\subsection{Energetics of Estimates}

Thermodynamic learning concerns agents that maximize work production from their
input data. As such, we now restrict our attention to agents that harness all
available nonequilibrium free energy in the form of work $\langle W \rangle =
-\Delta F^\text{neq}$. These \emph{maximum-work agents} zero out the average entropy
production $\langle \Sigma \rangle =-\langle W \rangle - \Delta F^\text{neq}$
and the work production of a computational map satisfies Eq.
(\ref{eq:EfficientWork2}). From here on out, when we refer to \emph{efficient}
agents we refer to those that maximize work production from the available
change in nonequilibrium free energy.

SOI state probabilities feature centrally in the expression for nonequilibrium
free energy and, thus, for the work production of efficient agents. However,
the \emph{actual} input distribution $\Pr(Z_\tau)$ may vary while the agent,
defined by its Hamiltonian $\mathcal{H}_\mathcal{Z}(t)$ over the computation
interval, remains fixed. Moreover, since the work production $\left\langle
W^\text{eff}_{|z_\tau,z_{\tau'}} \right\rangle$ of a computational map
explicitly conditions on the initial and final SOI state, this work cannot
explicitly depend on the input distribution. At first blush, this is a
contradiction: work that simultaneously does and does not depend on the input distribution.

This is resolved once one recognizes the role that estimates play in
thermodynamics. As indicated in Fig. \ref{fig:ThermodynamicLearning}, 
we claim that an agent has an estimated model of its environment that it uses to predict 
the SOI.  This model is, in one form or another, encoded in the evolving 
Hamiltonian $\mathcal{H}_\mathcal{Z}(t)$ that determines both the agent's 
energetic interactions and its logical architecture.  If an agent's estimated model 
of SOI $\mathcal{Z}$ is encoded as parameters $\theta$, then the agent estimates that the
SOI state $z$ at time $t$ has probability:
\begin{align*}
\Pr(Z^\theta_t=z_t)=\Pr(Z_t=z_t|\Theta=\theta)
  ~.
\end{align*}
The physical relevance of the estimated distribution comes from insisting that
\emph{the agent dissipates as little work as possible from a SOI whose
distribution matches its own estimate.}  In essence, the initial estimated
distribution $\Pr(Z_\tau^\theta)$ must be one of the distributions that
minimizes the average entropy production \cite{Kolc17a}:
\begin{align*}
\Pr(Z^\theta_\tau) \in \underset{\Pr(Z_\tau)}{\text{arg min}} \langle \Sigma \left[ \Pr(Z_\tau)\right] \rangle.
\end{align*}
Estimated probabilities $\Pr(Z^\theta_t)$ at later times $t>\tau$ are determined by updating the initial estimate via the stochastic dynamics that result from the Hamiltonian $\mathcal{H}_\mathcal{Z}(t)$ interacting with the thermal bath.

Thus, since an efficient agent produces zero entropy when the SOI follows the
minimum dissipation distribution $\Pr(Z^\theta_t)$, the work it produces from a
computational map is:
\begin{align}
\left\langle W^\theta_{|z_\tau,z_{\tau'}} \right\rangle
  = -\Delta E_\mathcal{Z}
  + \kB T \ln \frac{\Pr(Z^\theta_{\tau}=z_{\tau})}{\Pr(Z^\theta_{\tau'}=z_{\tau'})}
  ~.
\label{eq:ModelWork}
\end{align}
In this, we replaced the superscript ``eff'' with ``$\theta$'' to emphasize that
the agent is designed to be thermodynamically efficient for that particular
estimated model. Specifying the estimated model is essential, since
misestimating the input distribution leads to dissipation and entropy
production \cite{Kolc17a, Riec20a}. Returning to thermodynamic learning, this
is how the model $\theta$ factors into the ratchet's operation: estimated
distributions explicitly determine the work production of computational
maps.

Appendix \ref{app:Thermodynamically Efficient Markov Channels} gives a concrete
quasistatic mechanism for implementing \emph{any} computation $M$ and achieving
the work given by Eq. (\ref{eq:ModelWork}). This directly demonstrates how the
model $\theta$ is built into the evolving energy landscape
$\mathcal{H}_\mathcal{Z}(t)$ that implements $M$. The model $\theta$ determines
the initial and final change in state energies: $\Delta E(z,\tau)=-k_B \ln
\Pr(Z^\theta_\tau=z)$ and $\Delta E(z,\tau')=k_B \ln \Pr(Z^\theta_{\tau'}=z)$.
This quasistatic protocol operates and produces the same work for a particular
input-output pair regardless of the \emph{actual} input distribution.

Since we focus on the energetic benefits derived from information itself rather
than those from changing energy levels, the example implementations we use also
start and end with the same flat energy landscape. Restricting to such
information-driven agents, we consider cases where $\Delta E_\mathcal{Z} = 0$,
whereby:
\begin{align}
\left\langle W^\theta_{|z_\tau,z_{\tau'}} \right\rangle
  = \kB T \ln \frac{\Pr(Z^\theta_{\tau}=z_{\tau})}{\Pr(Z^\theta_{\tau'}=z_{\tau'})}
  ~.
\label{eq:ModelWork2}
\end{align}
This provides a direct relationship between the work production $\langle
W^\theta_{|z_\tau,z_{\tau'}} \rangle$ from particular data realizations and
the model $\theta$ that the agent uses, via the estimates $\Pr(Z^\theta_t)$
provided by that model. This is an essential step in determining a model
through thermodynamic learning.

\subsection{Thermally Efficient Agents}

With the work production of a maximally-efficient computational map
established, we are poised to determine the work production for
thermodynamically-efficient agents. Specifically, consider an agent
parameterized by its logical architecture $M$ and model parameters $\theta$. As
described by the agent architecture in Sec. \ref{sec:Work Production of
Thermodynamic Agents}, the agent uses its memory $X_j$ to map inputs $Y_j$ to
outputs $Y'_j$ and to update its memory to $X_{j+1}$. In stochastic mapping the
SOI $\mathcal{Z}=\mathcal{X}\times \mathcal{Y}_j$ the model parameter $\theta$
provides an estimate of the distribution over the current initial state
$(X_{j}^\theta, Y_{j}^\theta)$ as well as the final state $(X^\theta_{j+1},
Y'^\theta_{j})$. Assuming the agent's logical architecture $M$ is executed
optimally, direct application of Eq. (\ref{eq:ModelWork2}) then says that the
expected work production of the computational map $x_jy_j \rightarrow
x_{j+1}y'_j$ is:
\begin{align}
\label{eq:AgentWork1}
\left\langle W^\theta_{j,x_jy_j \rightarrow x_{j+1}y'_j} \right\rangle \equiv \kB T \ln \frac{\Pr(X^\theta_j=x,Y^\theta_j=y)}{\Pr(X^\theta_{j+1}=x',Y^{'\theta}_j=y')}
  ~.
\end{align}
In this, the estimated final distribution comes from the logical architecture
updating the initial distribution:
\begin{align*}
\Pr(X^\theta_{j+1} =x',& Y^{'\theta}_j=y') \\
  &=\sum_{x,y}\Pr(X^\theta_j=x,Y^\theta_j=y) M_{xy \rightarrow x'y'}
  ~.
\end{align*}

Equation (\ref{eq:AgentWork1})'s expression for work establishes that all
functional aspects (logical operation and energetics) of an efficient agent are
determined by two factors:
\begin{enumerate}
      \setlength{\topsep}{-2pt}
      \setlength{\itemsep}{-4pt}
      \setlength{\parsep}{-2pt}
\item The logical architecture $M$ that specifies how the agent manipulates
	inputs and updates its own memory.
\item The estimated input distribution $\Pr(X_{j}^\theta, Y_{j}^\theta)$ for
	which the agents' execution of $M$ is optimized to minimize dissipation.
\end{enumerate}
Thus, we define a \emph{thermally-efficient agent} by the ordered pair $\{M, \Pr(X_{j}^\theta, Y_{j}^\theta)\}$.

So defined, we can calculate the work produced when such agents act on a
particular input sequence $y_{0:L}$. This is done by first considering the work
production of a particular sequence of agent memory states $x_{0:L+1}$ and
outputs $y_{0:L}'$:
\begin{align*}
\left\langle W^\theta_{|y_{0:L},y_{0:L}',x_{0:L+1}} \right\rangle
  & = \sum_{j=0}^{L-1}
  \langle W^\theta_{j, x_jy_j \rightarrow x_{j+1}y'_j} \rangle \\
  & = \kB T \ln \prod_{j=0}^{L-1}
  \frac{\Pr(X^\theta_j \! = \! x_j,Y^\theta_j \! = \! y_j)}
  {\Pr(X^\theta_{j+1} \! = \! x_{j+1},Y^{'\theta}_j \! = \! y'_j)}
  ~.
\end{align*}
Then, to obtain the average work produced from a particular input sequence
$y_{0:L}$, we average over all possible hidden-state sequences and $x_{0:L+1}$ and output sequences $y'_{0:L}$:
\begin{widetext}
\begin{align}
\label{eq:TransducerWork}
\left\langle W^\theta_{|y_{0:L}} \right\rangle
  & = \kB T \sum_{x_{0:L+1},y_{0:L}'}
  \Pr(Y_{0:L}'=y_{0:L}',X_{0:L+1}=x_{0:L+1}|Y_{0:L}=y_{0:L}) \left\langle W^\theta_{|y_{0:L},y_{0:L}',x_{0:L+1}} \right\rangle \\
  \nonumber
  & = \kB T \sum_{x_{0:L+1},y_{0:L}'} \Pr(X_0=x_0)
  \prod_{k=0}^{L-1}M_{x_k,y_k \rightarrow x_{k+1},y_k'}   \ln \prod_{j=0}^{L-1}
  \frac{\Pr(X^\theta_j=x_j,Y^\theta_j=y_j)}
  {\Pr(X^\theta_{j+1}=x_{j+1},Y^{'\theta}_j=y'_j)}
  ~.
\end{align}
\end{widetext}
This gives the average energy harvested by an agent that transduces inputs
$y_{0:L}$ according to the logical architecture $M$, given that it is designed
to be as efficient as possible when its model $\theta$ matches the environment. 

On its own, Eq. (\ref{eq:TransducerWork})'s work production is a deeply
interesting quantity. In point of fact, since our agents are stochastic Turing
machines \cite{Broo89a}, this is the work production for any general form of
computation that maps inputs to output distributions
$\Pr(Y'_{0:L}|Y_{0:L}=y_{0:L})$ \cite{Barn13a}. Thus, Eq.
(\ref{eq:TransducerWork}) determines the possible \emph{work benefit for
universal thermodynamic computing}.

Given this general expression for work production, one might conclude that the
next step for thermodynamic learning is to search for the agent tuple $\{M,
\Pr(X_{j}^\theta, Y_{j}^\theta)\}$ that maximizes the work production.
However, this strategy comes with two issues. First, it requires a wider search
than necessary. Second, it does not draw a direct connection to the underlying
model $\theta$. Recall that we are considering \eM\ models $\theta$ of the
input sequence that give the probability estimate $\Pr(Y^\theta_{0:L}=y_{0:L})$
for any $L$.

We address both of these issues by refining the search space to agents whose
anticipated inputs $\Pr(X_{j}^\theta, Y_{j}^\theta)$ are explicitly determined
by their initial state distribution $\Pr(X_0^\theta)$ and estimated input
process $\Pr(Y^\theta_{0:\infty})$:
\begin{align*}
\Pr(X^\theta_j & = x_j,Y^\theta_j=y_j) \\
  & = \sum_{x_{0:j},y_{0:j},y'_{0:j}}
  \Pr(X^\theta_0=x_0)\Pr(Y^\theta_{0:j+1}=y_{0:j+1}) \\
  & \qquad \qquad \times \prod_{k=0}^{j-1}M_{x_k,y_k \rightarrow x_{k+1},y_k'}
  .
\end{align*}
We use this estimate for the initial state in Eq. (\ref{eq:TransducerWork}),
since it is maximally efficient, dissipating as little as possible if the agent
architecture $M$ receives the input distribution $\Pr(Y^\theta_{0:\infty})$. As
a result of its efficiency, the resulting computation performed by the agent
produces the maximum possible work, given its logical architecture.

This simplifies our search for maximum-work agents by directly tying the
estimated inputs to the model $\theta$. However, it still leaves one piece of
the agent undefined: its logical architecture $M$. Fortunately, as we discuss
now, the thermodynamics of modularity further simplifies the search.

\subsection{Thermodynamics of Modularity}

An agent transduces inputs to outputs through a series of modular operations.
The Hamiltonian $\mathcal{H}_{\mathcal{X} \times \mathcal{Y}_j}(t)$ that
governs the evolution of the $j$th operation is decoupled from the other
elements of the time series $\mathcal{Y}_0 \times \mathcal{Y}_1 \cdots
\mathcal{Y}_{j-1} \times \mathcal{Y}_{j+1} \times \cdots$. As a result of this
modular computational architecture, the correlations lost between the agent and
the rest of the information reservoir are irreversibly dissipated, producing
entropy.  This is an energetic cost associated directly with the agent's
logical architecture, known as the \emph{modularity dissipation}
\cite{Boyd17a}. It sets a minimum for the work dissipated in a complex
computation composed of many elementary steps.

To continue our pursuit of maximum-work, we must design the agent's logical
architecture to minimize dissipated work. Past analysis of agents that
harness energy from a pattern $\Pr(Y_{0:\infty})$ showed that the modularity
dissipation is only minimized when the agent's states are \emph{predictive} of
the pattern \cite{Boyd17a}. This means that to maximize work extracted, an
agent's state must contain all information about the past relevant to the
future $\Pr(Y_{j+1:L}|X_j) = \Pr(Y_{j+1:L}|Y_{0:j})$. That is, for maximal work
extraction agent states must be sufficient statistics for predicting the
future.

Moreover, the \eMs\ introduced in Sec. \ref{sec:CMech} are constructed with
hidden states $S_j$ that are a \emph{minimal} predictor of their output
process. This is why they are referred to as \emph{causal states}. And
so, the \eM\ is a minimal sufficient statistic for prediction. Transitions
among causal states trigger outputs according to:
\begin{align*}
\theta^{(y)}_{s \rightarrow s'}=\Pr(Y^\theta_j=y,S^\theta_{j+1}=s'|S^\theta_j=s)
  ~,
\end{align*}
which is the probability that an \eM\ in internal state $s$ transitions to $s'$
and emits output $y$. \EMs\ have the additional property that they are unifilar,
meaning that the next causal state $s'$ is uniquely determined by the current
state $s$ and output $y$ via the propagator function
$s'=\epsilon(s,y)$.

In short, to produce maximum work agent memory must store at least the causal
states of the environment's own \eM. Appendix \ref{app:Designing Agents}
describes how the logical architecture of the maximum-work minimum-memory agent
is determined by its estimated \eM\ model $\theta$ of its inputs. The agent
states are chosen to be the same as the causal states
$\mathcal{X}=\mathcal{S}$, and they update according to the propagator function
$\epsilon$:
\begin{align}
M_{xy \rightarrow x'y'} \! = \! \frac{1}{|\mathcal{Y}_j|} \! \times \!
  \begin{cases}
  & \delta_{x',\epsilon(x,y)} \text{, if } \sum_{x'}\theta^{(y)}_{x \rightarrow x'} \neq 0, \\
  & \delta_{x',x} \text{, otherwise.}
\end{cases}
  ~.
\label{eq:Agents1FromModel}
\end{align}
This guarantees that the agent's internal states follow the causal states
of its estimated input process $\Pr(Y^\theta_{0:\infty})$. In turn, it prevents
the agent from dissipating temporal correlations within that process.

\subsection{Agent--Model Equivalence}
\label{Agent--Model Equivalence}

In the pursuit of maximum work, we refined the structure of candidate agents
considerably, limiting consideration to those that minimize the modularity dissipation. As a result, they store the predictive states of their estimated
input and their logical architecture is explicitly determined by an \eM\ model.
Moreover, to be efficient, the candidate agent should begin in the \eM's start
state $s^*$ such that the model $\theta$ uniquely determines the second piece
of an efficient agent. This is the estimated initial distribution
over its memory state and input:
\begin{align*}
\Pr(Y^\theta_j=y_j,X^\theta_j=s_j)
  = \sum_{y_{0:j},s_{0:j},s_{j+1}} \delta_{s_0,s^*}
  \prod_{k=0}^{j} \theta^{(y_k)}_{s_k \rightarrow s_{k+1}}
  ~.
\end{align*}

Conversely, maximum-work agents, characterized by their logical architecture
and anticipated input distributions $\{M,\Pr(X^\theta_j,Y^\theta_j)\}$, also
specify the \eM's model of their estimated distribution:
\begin{align}
\theta^{(y)}_{s \rightarrow s'}
  & = \Pr(Y^\theta_j=y|S^\theta_j=s) \delta_{s',\epsilon(s,y)} \nonumber \\
  & = \Pr(Y^\theta_j=y|X^\theta_j=s)|\mathcal{Y}_j|M_{sy \rightarrow s' y'}
  ~.
\label{eq:ModelFromAgent}
\end{align}
Through the \eM, the agent also specifies its estimated input process. In this
way, we arrive at a class of agents that  are uniquely determined by their
environment model.

\begin{figure*}[tbp]
\centering
\includegraphics[width=2\columnwidth]{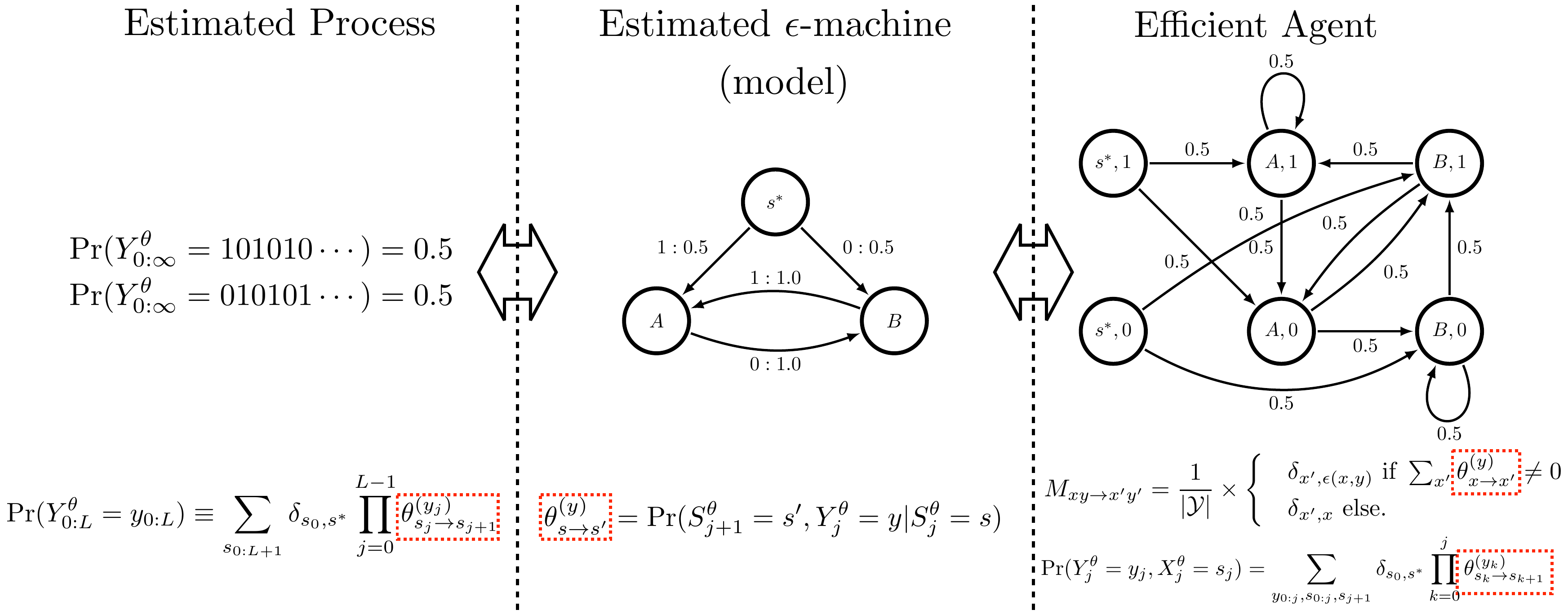}
\caption{Equivalence of estimated input process
	$\Pr(Y^\theta_{0:\infty}=y_{0:\infty})$, \eM\ $\theta$, and the agent that
	efficiently harnesses the input process asymptotically, using logical
	architecture $M_{xy \rightarrow x'
	y'}=\Pr(X_{j+1}=x',Y_{j+1}=y'|X_j=x,Y_j=y)$ and estimated input
	distribution $\Pr(X^\theta_j=x,Y^\theta_j=y)$. Determining one determines
	the others.
	}
\label{fig:AgentModelEquivalence}
\end{figure*}

Figure \ref{fig:AgentModelEquivalence} explicitly lays this out. It presents
an agent that estimates a period-$2$ process with uncertain phase, such that
$\Pr(Y^\theta_{0:\infty}=0101\cdots)=\Pr(Y^\theta_{0:\infty}=1010\cdots)=0.5$.
The middle column shows the \eM\ that is uniquely determined for that process,
characterized by the model parameters $\theta^{(y)}_{s \rightarrow s'}$. The
right column shows the unique minimal agent that harnesses as much work as
possible from that process. All three, (i) the estimated process, (ii) the
estimated $\epsilon$-machine model, and (iii) the maximum-work agent are
equivalent. And, they can be determined from one another. This holds true for
\emph{any} estimated input process $\Pr(Y^\theta_{0:\infty})$.

Under the equivalence of model $\theta$ and agent operation, when we monitor
the agent's thermodynamic performance through its work production, we also
measure the predictive performance of its underlying model.

This completes the thermodynamic learning framework laid out in Fig.
\ref{fig:ThermodynamicLearning}. There, the model an agent holds affects its
interaction with the symbol sequence $y_{0:L}$ and, ultimately, its work
production. And so, from this point forward, when discussing an estimated
process or an \eM\ that generates that guess, we are also describing the unique
thermodynamic agent designed to produce maximal work from the estimated
process. We can now turn to explore how such agents' work production ties to
their underlying models. A direct comparison to log-likelihood
parametric-density estimation can now be drawn.

\section{Work-Likelihood Correspondence for Agent Design}
\label{sec:Designing Agents}

We are now ready to return to our core objective---exploring work production as
a performance measure for a model estimated from a time series $y_{0:L}$. In
comparison to the expression for general computing in Eq.
(\ref{eq:TransducerWork}), using efficiently-designed predictive agents leads
to a much simpler expression for work production:
\begin{empheq}{align}
\left\langle W^\theta_{|y_{0:L}} \right\rangle
  \! = \! \kB T \left( \ln \Pr(Y^\theta_{0:L} \! = \! y_{0:L})
  \! + \! L \ln |\mathcal{Y}| \right)
  ~.
\end{empheq}
The mechanism behind this vast simplification arises from unifilarity---a
property of prediction machines that guarantees a single state trajectory on
$x_{0:L}$ for each input string $y_{0:L}$. The details of the derivation are
outlined in Appendix \ref{app:Work Production of Optimal Transducers}.

This expression directly captures the relationship between work production and
the agent's underlying model of the data. To see this, we recast it in the
language of machine learning. Consider $y_{0:L}$ as training data in
parametric density estimation. We are then tasked to construct a model of this
data. Each candidate model is parameterized by $\theta$, which results in an
estimated process $Y^\theta_{0:L}$. Observe then that $\Pr(Y^\theta_{0:L} \! =
\! y_{0:L})$ is simply the probability that the candidate model will output
$y_{0:L}$. Therefore, the log-likelihood $\ell(\theta|y_{0:L})$ of parametric
estimation coincides with $\ln \Pr (Y^\theta_{0:L} = y_{0:L})$, and we can
write:
\begin{empheq}[box=\fbox]{align}
\label{eq:AgentWork}
\left\langle W^\theta_{|y_{0:L}} \right\rangle
  & = \kB T \ell(\theta|y_{0:L})+ \kB T L \ln |\mathcal{Y}|
  ~.
\end{empheq}
One concludes that work production is maximized precisely when the
log-likelihood $\ell(\theta|y_{0:L})$ is maximized. Thus, the criterion for
\emph{creating a good model of an environment} is the same as that for
\emph{extracting maximal work}.

This link is made concrete via the simple example presented in App.
\ref{sec:Training Memoryless Agents}. It goes through an explicit description of
the Hamiltonian control required to implement a memoryless agent that harvests
work from a sequence of up spins $\uparrow$ and down spins $\downarrow$ that
compose the time series $y_{0:L}$. The agent's internal memoryless model
results in Eq. (\ref{eq:AgentWork})'s work production. And, we
find that the maximum-work agent has learned about the input sequence.
Specifically, the agent learns the frequency of spins $\uparrow$ and
$\downarrow$, confirming the basic principle of maximum-work thermodynamic
learning. However, the learning presented in App. \ref{sec:Training
Memoryless Agents} precludes the possibility of learning temporal structure in
the spin sequence, since the agents and their internal models have no memory
\cite{Boyd16d}. To learn about the temporal correlations within the sequence,
one must use agents with multiple memory states. We leave thermodynamic
learning among memoryful agents for later investigation.

Stepping back, we see the relationship between machine learning and information thermodynamics more clearly. In parametric density estimation we have:
\begin{enumerate}
      \setlength{\topsep}{-2pt}
      \setlength{\itemsep}{-4pt}
      \setlength{\parsep}{-2pt}
\item Data $y_{0:L}$ that provides a window into a black box.
\item A model $\theta$ of the black box that determines an estimated
	distribution over the data $\Pr(Y^\theta_{0:L})$.
\item A performance measure for the model of the data, given by the
	log-likelihood $\ell (\theta|y_{0:L})= \ln \Pr(Y^\theta_{0:L}=y_{0:L})$.
\end{enumerate}
The parallel in thermodynamic learning is exact, with:
\begin{enumerate}
\item Data $y_{0:L}$ physically stored in systems
	$\mathcal{Y}_0 \times \mathcal{Y}_1 \times \dots \mathcal{Y}_{L-1}$
	output from the black box.
\item An agent $\{M,\Pr(X^\theta_j,Y^\theta_j)\}$ that is entirely determined
	by the model $\theta$.
\item The agent's thermodynamic performance, given by its work production
	$\left \langle W^\theta_{|y_{0:L}} \right\rangle $, increases linearly
	with the log-likelihood $\ell (\theta|y_{0:L})$.
\end{enumerate}
In this way, we see that thermodynamic learning through work maximization is
\emph{equivalent} to parametric density estimation. 

Intuitively, the natural world is replete with complex learning systems---an
observation seemingly at odds with Thermodynamics and its Second Law which
dictates that order inevitably decays into disorder. However, our results are
tantamount to a contravening physical principle that drives the emergence of
order through learning: work maximization. We showed, in point of fact, that
work maximization and learning are equivalent processes. At a larger remove,
this hints of general physical principles of emergent organization.

\section{Searching for Principles of Organization}
\label{sec:PrincOrg}

Introducing an \emph{equivalence of maximum work production and optimal
learning} comes at a late stage of a long line of inquiry into what kinds of
thermodynamic constraints and laws govern the emergence of organization and,
for that matter, biological life. So, let's historically place the
seemingly-new principle. In fact, it enters a crowded field.

Within statistical physics the paradigmatic principle of organization was found
by Kirchhoff \cite{Kirc48a}: in electrical networks current distributes itself
so as to dissipate the least possible heat for the given applied voltages.
Generalizations, for equilibrium states, are then found in Gibbs' variational
principle for entropy for heterogeneous equilibrium \cite{Gibb06a}, Maxwell's
principles of minimum-heat \cite[pp. 407-408]{Maxw54a}, and Onsager's
minimizing the ``rate of dissipation'' \cite{Onsa31a}.

Close to equilibrium, Prigogine introduced minimum entropy production
\cite{Prig45a}, identifying \emph{dissipative structures} whose
maintenance requires energy \cite{Prig71b}. However, far from
equilibrium the guiding principles can be quite the opposite. And so, the
effort continues today, for example, with recent applications of nonequilibrium
thermodynamics to pattern formation in chemical reactions \cite{Fala18a}. That
said, statistical physics misses at least two, related, but key components:
dynamics of and information in thermal states.

Dynamical systems theory takes a decidedly mechanistic approach to the
emergence of organization, analyzing the geometric structures in a system's
state space that amplify fluctuations and eventually attenuate them into
macroscopic behaviors and patterns. This was eventually articulated by pattern
formation theory \cite{Turi52,Hoyl06a,Cros09a}. A canonical example is
fluid turbulence \cite{Heis67a}---a dynamical explanation for its complex
organizations occupied much of the 70s and 80s. Landau's original theory of
incommensurate oscillations was superseded by the mathematical discovery in the
1950s of chaotic attractors \cite{Ruel71a,Bran83}. This approach, too, falls
short of leading to a principle of emergent organization. Patterns emerge, but
what exactly are they and what complex behavior do they exhibit?

Answers to this challenge came from a decidedly different direction---Shannon's
theory of noisy communication channels and his measures of information
\cite{Shan48a,Cove06a}, appropriately extended \cite{Jame11a}. While
adding an important new perspective---that organized systems store and transmit
information---this, also, did not go far enough as it side-stepped the content
and meaning of information \cite{Shan56b}. In-roads to these appeared in the
theory of computation inaugurated by Turing \cite{Turi37a}. The most direct and
ambitious approach to the role of information in organization, though, appeared
in Wiener's cybernetics \cite{Wien48,Ashb57a}. While it eloquently laid out the
goals to which principles should strive, it ultimately never harnessed the
mathematical foundations and calculational tools needed. Likely, the earliest
overt connection between statistical mechanics and information, though,
appeared with Jaynes' Maximum Entropy \cite{Jayn57a} and Minimum
Entropy Production Principles \cite{Jayn80a}.

So, what is new today is the synthesis of statistical physics, dynamics, and
information. This, finally, allows one to answer the question, How do physical
systems store and process information? The answer is that they
\emph{intrinsically compute} \cite{Crut12a}. With this, one can extract from
behavior a system's information processing, even going so far as to discover
the effective equations of motion \cite{Kolm58,Pack80,Take81,Crut87a}. One can
now frame questions about how a physical system reacts to, controls, and adapts
to its environment.

All such systems, however, are embedded in the physical world and require
resources to operate. More to the point, what energetic resources underlie
computation? Initiated by Brillouin \cite{Bril62a} and Landauer and Bennett
\cite{Land61a,Benn82}, today there is a nascent \emph{physics of information}
\cite{Saga13,Parr15a}. Resource constraints on computing by thermodynamic
systems are now expressed in a suite of new principles. For example, the
\emph{principle of requisite complexity} \cite{Boyd16d} dictates that
maximally-efficient interactions require an agent's internal organization match
the environment's organization. And, thermodynamic resource costs arise from
the modularity of an agent's architecture \cite{Boyd17a}. Pushing the search
for organization further, the preceding established the thermodynamics of how a
system learns, suggesting the possibility of adaptive organization.

To fully appreciate organization in natural processes, though, one must also
address dynamics of agent populations, first on the time scale of agent life
cycles and second on the scale of many generations. In fact, tracking the
complexity of individuals reveals that selection pressures spontaneously emerge
in purely-replicating populations \cite{Crut04a} and replication itself
necessarily dissipates energy \cite{Engl13a}.

As these pieces assembled, a picture has come into focus. Intelligent, adaptive
systems learn to harness resources from their environment, expending energy to
live and reproduce. Taken altogether, the historical perspective suggests we are moving close to realizing Wiener's cybernetics \cite{Wien48}.

\section{Conclusion}

We introduced thermodynamic machine learning---a physical process that trains
intelligent agents by maximizing work production from complex environmental
stimuli supplied as time-series data. This involved constructing a framework to
describe thermodynamics of computation at the single-shot level, enabling us to
evaluate the work an agent can produce from individual data realizations. Key
to the framework is its generality---applicable to agents exhibiting arbitrary
adaptive input-output behavior and implemented within any physical substrate.

In the pursuit of maximum work, we refined this general class of agents to those
that are best able to harness work from temporally-correlated inputs. We found
that the performance of such maximum-work agents increases proportionally to
the log-likelihood of the model they use for predicting their environment. As
a consequence, our results show that thermodynamic learning exactly mimics
parametric density estimation in machine learning. Thus, work is a
thermodynamic performance measure for physically-embedded learning. This result
further solidifies the connections between agency, intelligence, and the
thermodynamics of information---hinting that energy harvesting and learning may
be two sides of the same coin.

These connections suggest a number of exciting future directions. From the
technological perspective, they hint at a natural method for designing
intelligent energy harvesters---establishing that our present tools of machine
learning can be directly mapped to automated design of efficient information
ratchets and pattern engines~\cite{Serr07a,Garn15,Boyd16d}. Meanwhile, recent
results indicate that quantum systems can generate complex adaptive behaviors
using fewer resources than classical
counterparts~\cite{Thom17a,Loom19a,Wood18a}. Does this suggest there are new
classes of quantum-enhanced energy harvesters and learners?

Ultimately, energy is an essential currency for life. This highlights the
question, To what extent is work optimization a natural tendency of driven
physical systems? Indeed, recent results indicate physical systems evolve to
increase work production \cite{Gold19a, Zhon20a}, opening a fascinating
possibility. Could the equivalence between work production and learning then
indicate that the universe itself naturally learns?  The fact that complex intelligent life emerged from the lifeless soup of the universe might be
considered a continuing miracle: a string of unfathomable statistical anomalies
strung together over eons. It would certainly be extraordinary if this evolution
then has a physical basis---hidden laws of thermodynamic organization that guide
the universe to create entities capable of extracting maximal work.

\section*{Acknowledgments}

The authors thank the Telluride Science Research Center for hospitality during
visits and the participants of the Information Engines Workshops there.  ABB
thanks Wesley Boyd for useful conversations and JPC similarly thanks Adam Rupe.
JPC acknowledges the kind hospitality of the Santa Fe Institute, Institute for
Advanced Study at the University of Amsterdam, and California Institute of
Technology for their hospitality during visits. This material is based upon
work supported by, or in part by, Grant No. FQXi-RFP-IPW-1902 and FQXi-RFP-1809
from the Foundational Questions Institute and Fetzer Franklin Fund (a
donor-advised fund of Silicon Valley Community Foundation), the Templeton World
Charity Foundation Power of Information fellowship TWCF0337 and TWCF0560, the
National Research Foundation, Singapore, under its NRFF Fellow program (Award
No. NRF-NRFF2016-02), Singapore Ministry of Education Tier 1 Grants No.
MOE2017-T1-002-043, and U.S. Army Research Laboratory and the U.S. Army
Research Office under grants W911NF-18-1-0028 and W911NF-21-1-0048. Any
opinions, findings and conclusions or recommendations expressed in this
material are those of the authors and do not reflect the views of National
Research Foundation, Singapore.

\emph{During submission we became aware of related work: L. Touzo, M. Marsili,
N. Merhav, and E. Roldan.} Optimal work extraction and the minimum description
length principle. \emph{arXiv:2006.04544.}

\appendix

\section{Extended Background}
\label{app:Extended Background}

Developing the principle of maximum work production and calling out the
physical benefits of an agent modeling its environment drew heavily from the
areas of computational mechanics, nonequilibrium thermodynamics, and machine
learning. Due to the variety of topics addressed, the following provides a more
detailed notational and conceptual summary. This should aid the development be
more self-contained, hopefully providing a common language across the areas and
a foundation for further exploration. While we make suggestive comparisons by
viewing foundations of each area side-by-side, it may be most appropriate for
readers already familiar with them and concerned with novel results to skip,
using the review to clarify unfamiliar notation.

\subsection{Machine Learning and Generative Models}

Thermodynamically, what is a good model of the data with which an agent
interacts? Denote the data's state space as $\mathcal{Z}=\{z \}$. If we
have many copies of $\mathcal{Z}$, all initially prepared in the same way, then
as we observe successive realizations $\vec{z}=\{z_0,z_1, \cdots, z_N\}$ from
an ensemble, the frequency of an observed state $z$ approaches the
\emph{actual} probability distribution $\Pr(Z=z)$, where $Z$ is the random
variable that realizes states $z \in \mathcal{Z}$. However, with only a finite
number $N$ of realizations, the best that can be done is to characterize the
environment with an \emph{estimated} distribution $\Pr(Z^\theta=z)$. Estimating
models that agree with finite data is the domain of statistical inference and
machine learning of generative models \cite{Shai14a, Reze14a, Daya95a}.

At first blush, estimating a probability distribution appears distinct from
familiar machine learning challenges, such as image classification and and the
inverse problem of artificially generating exemplar images from given data.
However, both classification and prediction can be achieved through a form of
unsupervised learning \cite{Lin17a}. For instance, if the system is a joint
variable over both the pixel images and the corresponding label
$\mathcal{Z}=\text{pixels}\times\{\text{cat}, \text{dog}\}$, then our estimated
distribution $\Pr(Z^\theta=z)$ gives both a means of choosing a label for an
image $\Pr(\text{label}^\theta=\text{cat}|\text{pixels}^\theta=\text{image})$
and a means of choosing an image for a label
$\Pr(\text{pixels}^\theta=\text{image}|\text{label}^\theta=\text{cat})$.

A generative model is specified by a set of parameters $\theta$ from which the
model produces the estimated distribution
$\Pr(Z^\theta=z)=\Pr(Z=z|\Theta=\theta)$. The procedure of arriving at this
estimated model is parametric density estimation \cite{Shai14a,Daya95a}.
However, we take the random variable $Z^\theta$ for the estimated distribution
to denote the model for notational and conceptual convenience.

The Shannon entropy \cite{Cove06a}:
\begin{align*}
H[Z]\equiv -\sum_z\Pr(Z=z) \ln \Pr(Z=z)
\end{align*}
measures uncertainty in nats, a ``natural'' unit for thermodynamic entropies.
The Shannon entropy easily extends to joint probabilities and all information
measures that come from their composition (conditional and mutual
informations). For instance, if the environment is composed of two correlated
subcomponents $\mathcal{Z}=\mathcal{X} \times \mathcal{Y}$, the probability and
entropy are expressed:
\begin{align*}
\Pr(Z=z) & = \Pr(X=x,Y=y) \\
   & \neq \Pr(X=x)\Pr(Y=Y) ~\text{and}\\
   H[Z] & = H[X,Y] \\
   & \neq H[X]+H[Y]
  ~,
\end{align*}
respectively.

While there are many other ways to create parametric models $\theta$---from
polynomial functions with a small number of parameters to neural networks
with thousands \cite{Shai14a}---the goal is to match as well as possible the
estimated distribution $\Pr(Z^\theta)$ to the actual distribution $\Pr(Z)$.

One measure of success in this is the probability that the model generated the
data---the \emph{likelihood}. The likelihood of the model $\theta$ given a data
point $z$ is the same as the likelihood of $Z^\theta$:
\begin{align*}
\mathcal{L}(\theta|z) & = \Pr(Z=z|\Theta=\theta) \\
  & =\Pr(Z^\theta=z) \\
  & =\mathcal{L}(Z^\theta|z)
  ~.
\end{align*}

Given a set $\vec{z}=\{z_1,z_2, \cdots,  z_N\}$ of training data and assuming
independent samples, then the likelihood of the model is the product:
\begin{align}
\mathcal{L}(Z^\theta|\vec{z})= \prod_{i=1}^N \mathcal{L}(Z^\theta|z_i)
  ~.
\end{align}
This is a commonly used performance measure in machine learning, where
algorithms search for models with maximum likelihood \cite{Shai14a}. However,
it is common to use the \emph{log-likelihood} instead, which is maximized for
the same models:
\begin{align}
\ell(\theta|\vec{z}) & = \ln \mathcal{L}(\theta|\vec{z}) \nonumber \\
  & =  \sum_{i=1}^N \ln \Pr(Z^\theta=z_i)
  ~.
\label{eq:loglikelihood}
\end{align}

If the model $Z^\theta$ were specified by a neural network, the log-likelihood
could be determined through stochastic gradient descent back-propagation
\cite{Reze14a, LeCu15a}, for instance. The intention is that the procedure will
converge on a network model that produces the data with high probability.

\subsection{Thermodynamics of Information}

Learning from data translates information in an environment into a useful
model. What makes that model useful? In a physical setting, recalling from
Landauer that ``information is physical'' \cite{Land91a}, the usefulness one
can extract from thermodynamic processes is work. Figure
\ref{fig:PhysicalComputation} shows a basic implementation for physical
computation. Such an information-storing physical system $\mathcal{Z}=\{z\}$,
in contact with a thermal reservoir, can execute useful computations by drawing
energy from a work reservoir. Energy flowing from the system $\mathcal{Z}$ into
the thermal reservoir is positive heat $Q$. When energy flows from the system
$\mathcal{Z}$ to the work reservoir, it is positive work $W$ production.  Work
production quantifies the amount of energy that is stored in the work reservoir
available for later use. And so, in this telling, it represents a natural and
physically-motivated measure of thermodynamic performance. In the framework for
thermodynamic computation of Fig.  \ref{fig:PhysicalComputation}, work is
extracted via controlling the system's Hamiltonian.

Specifically, the system's informational states are controlled via a
time-dependent Hamiltonian---energy $E(z,t)$ of state $z$ at time $t$. For
state trajectory $z_{\tau:\tau'}=z_\tau z_{\tau+dt} \cdots z_{\tau'-dt}
z_{\tau'}$ over time interval $t \in [\tau, \tau']$, the resulting work
extracted by Hamiltonian control is the temporally-integrated change in energy
\cite{Deff2013}:
\begin{align*}
 W_{|z_{\tau:\tau'}}  =- \int_{\tau}^{\tau'}
 	dt\, \partial_t E(z,t)\big\vert_{z=z_t}
 ~.
\end{align*}

Heat $Q_{|z_{\tau:\tau'}}=E(z_{\tau},\tau)-E(z_{\tau'},\tau')-W_{|z_{\tau:\tau'}}$ flows into the thermal reservoir, increasing its entropy:
\begin{align}
\Delta S^\text{reservoir}_{|z_{\tau:\tau'}}= \frac{Q_{|z_{\tau:\tau'}}}{T}
  ~.
\end{align}
where the thermal reservoir is at temperature $T$. The Second Law of
Thermodynamics states that, on average, any processing on the informational
states can only yield nonnegative entropy production of the universe (reservoir
and system $\mathcal{Z}$):
\begin{align}
\langle \Sigma \rangle
  & = \langle \Delta S^\text{reservoir} \rangle
  + \langle \Delta S^\mathcal{Z} \rangle \nonumber \\
  & \geq 0
  ~.
\label{eq:SecondLaw}
\end{align}
This constrains the energetic cost of computations performed within the system $\mathcal{Z}$.

A computation over time interval $t \in [\tau,\tau']$ has two components:
\begin{enumerate}
\item An initial distribution over states $\Pr(Z_\tau=z_\tau)$, where $Z_t$
	is the random variable of system $\mathcal{Z}$ at time $t$.
\item A Markov channel that transforms it, specified by the conditional
	probability of the final state $z_{\tau'}$ given the initial input
	$z_\tau$:
\begin{align*}
M_{z_\tau \rightarrow z_{\tau'}}= \Pr(Z_{\tau'}=z_{\tau'}|Z_\tau=z_\tau)
  ~.
\end{align*}
\end{enumerate}

This specifies, in turn, the final distribution $\Pr(Z_{\tau'}=z_{\tau'})$
that allows direct calculation of the system-entropy change \cite{Espo11}:
\begin{align*}
\Delta S^\mathcal{Z}_{|z_{\tau:\tau'}}
  = \kB \ln \frac{\Pr(Z_\tau=z_\tau)}{\Pr(Z_{\tau'}=z_{\tau'})}
  ~.
\end{align*}
Adding this to the information reservoir's entropy change yields the entropy
production of the universe. This can also be expressed in terms of the work
production:
\begin{align*}
\Sigma_{|z_{\tau:\tau'}}
  & \equiv \Delta S^\text{reservoir}_{|z_{\tau:\tau'}}
  + \Delta S^\mathcal{Z}_{|z_{\tau:\tau'}} \\
  & = \frac{-W_{|z_{\tau:\tau'}}+\phi(z_\tau,\tau)-\phi(z_{\tau'},\tau')}{T}
  ~.
\end{align*}
Here, $\phi(z,t)=E(z,t)+ \kB T \ln \Pr(Z_t=z) $ is the \emph{pointwise
nonequilibrium free energy}, which becomes the \emph{nonequilibrium free
energy} when averaged: $\langle \phi(z,t) \rangle_{\Pr(Z_t=z)}=F^\text{neq}(t)$
\cite{Deff12a}.

Note that the entropy production is also proportional to the additional work
that could have been extracted if the computation was efficient. This is
referred to as the \emph{dissipated work}:
\begin{align}
W^\text{diss}_{|z_\tau:z_{\tau'}} = T \Sigma_{|z_\tau:z_{\tau'}}
  ~.
\end{align}

Turning back to the Second Law of Thermodynamics, we see that the
\emph{average} work extracted is bounded by the change in nonequilibrium free
energy:
\begin{align*}
\langle W\rangle \geq F^\text{neq}(\tau')-F^\text{neq}(\tau)
  ~.
\end{align*}
When the system starts and ends as an information reservoir, with equal
energies for all states $E(z,\tau)=E(z',\tau')$ \cite{Deff2013}, this reduces
to Landauer's familiar principle for erasure \cite{Land61a}---work production
must not exceed the change in state uncertainty:
\begin{align*}
\langle W\rangle \leq \kB T (H[Z_{\tau'}]-H[Z_{\tau}])
  ~,
\end{align*}
where $H[Z_t]=-\sum_{z \in \mathcal{Z}}\Pr(Z_t=z) \ln \Pr(Z_t=z)$ is the
Shannon entropy of the system at time $t$ measured in nats. This is the
starting point for determining the work production that agents can extract from
data.

\subsection{Computational Mechanics}

When describing thermodynamics and machine learning, data was taken from the
state space $\mathcal{Z}$ all at once. However, what if we consider a state
space composed of $L$ identical components $\mathcal{Z}=\mathcal{Y}^L$ that are
received in sequence. Our model of the time series $y_{0:L}$ of realizations is
described by an estimated distribution $\Pr(Y^\theta_{0:L} = y_{0:L})$.
However, for $L$ large enough, this object becomes impossible to store, due to
the exponential increase in the number of sequences. Fortunately, there
are ways to generally characterize an arbitrarily long time-series distribution
using a finite model.

\subsubsection{Generative Machines}

A hidden Markov model (HMM) is described by a set of hidden states
$\mathcal{S}$, a set of output states $\mathcal{Y}$, a conditional
output-labeled matrix that gives the transition probabilities between the
states:
\begin{align*}
\theta^{(y)}_{s \rightarrow s'}=\Pr(S^\theta_{j+1}=s',Y^\theta_j=y|S^\theta_j=s)
\end{align*}
for all $j$, and a start state $s^* \in \mathcal{S}$. (Generally, one also
specifies an initial state distribution, with selecting a start state being a
special case.) We label the transition probabilities with the model parameter
$\theta$, since these are the actual parameters that must be stored to generate
probabilities of time series. For instance, Fig. \ref{fig:HMM} shows an HMM
that generates a periodic process with uncertain phase. Edges between hidden
states $s$ and $s'$ are labeled $y:\theta^{(y)}_{s \rightarrow s'}$, where $y$
is the output symbol and $\theta^{(y)}_{s \rightarrow s'}$ is the probability
of emitting that symbol on that transition.

If $\{ \theta^{(y)}_{s \rightarrow s'} \}$ is the model for the estimated input
$\Pr(Y^\theta=y_{0:L})$, then the probability of any word is calculated by
taking the product of transition matrices and summing over internal states:
\begin{align*}
\Pr(Y^\theta_{0:L}=y_{0:L})
  = \sum_{s_{0:L+1}}\delta_{s_0,s^*}
  \prod_{j=0}^{L-1} \theta^{(y_j)}_{s_j \rightarrow s_{j+1}}
  ~.
\end{align*}

Beyond generating length-$L$ symbol strings, these HMMs generate distributions
over semi-infinite strings $\Pr(Y^\theta_{0:\infty}=y_{0:\infty})$. As such,
they allow us to anticipate more than just the first $L$ symbols from the same
source. Once we have a model $\theta=\{(\theta_{s \rightarrow
s'}^{(y)},s,s',y)\}_{s,s',y}$ from our training data $y_{0:L}$, we can
calculate probabilities of longer words $\Pr(Y^\theta_{0:L'}=y_{0:L'}')$ and,
thus, the probability of symbols following the training data
$\Pr(Y^\theta_{L:L'}=y_{L:L'}'|Y^\theta_{0:L}=y_{0:L})$.

Distributions over semi-infinite strings $\Pr(Y^\theta_{0:\infty}=y_{0:\infty})$
are similar to processes, which are distributions over bi-infinite strings
$\Pr(Y^\theta_{-\infty:\infty}=y_{-\infty:\infty})$. While not insisting on
stationarity, and so allowing for flexibility in using subwords of length $L$,
we can mirror computational mechanics' construction of unifilar HMMs
from time series, where the hidden states $S^\theta_j$ are minimal sufficient
statistics of the past $Y^\theta_{0:j}$ about the future $Y^\theta_{j:\infty}$
\cite{Crut01a}. In other words, the hidden states are perfect predictors.

Given a semi-infinite process $\Pr(Y^\theta_{0:\infty}=y_{0:\infty})$, we
construct a minimal predictor through a causal equivalence relation $y_{0:k}
\sim y_{0:j}'$ that says two histories $y_{0:k}$ and $y_{0:j}'$ are members of
the same equivalence class if and only if they have the same semi-infinite
future distribution:
\begin{align*}
\Pr(Y^\theta_{j:\infty} \! = \! y_{0:\infty}''|Y^\theta_{0:j} \! = \! y_{0:j}')
  = \Pr(Y^\theta_{k:\infty} \! = \! y_{0:\infty}''|Y^\theta_{0:k} \! = \! y_{0:k})
  ~.
\end{align*}
An equivalence class of histories is a \emph{causal state}. Causal states also
induce a map $\epsilon(\cdot)$ from histories $y_{0:k}$ to states $s_i$:
\begin{align*}
s_i & =\{y_{0:j}'|y_{0:k} \sim y_{0:j}' \} \\
    & \equiv \epsilon(y_{0:k})
	~.
\end{align*}
This guarantees that a causal state is a sufficient statistic of the past about
the future, such that we can track it as a perfect predictor:
\begin{align*}
\Pr(Y^\theta_{k:\infty}|Y^\theta_{0:k}=y_{0:k})
  = \Pr(Y^\theta_{k:\infty}|S^\theta_k=\epsilon(y_{0:k}))
  ~.
\end{align*}
In fact, the causal states are minimal sufficient statistics.

Constructing causal states reveals a number of properties of stochastic
processes and models. One of these is \emph{unifilarity}, which means that if
the current causal state $s_k=\epsilon(y_{0:k})$ is followed by any sequence
$y_{k:j}$, then the resulting causal state $s_k=\epsilon(y_{0:k})$ is uniquely
determined.

And, we can expand our use of the $\epsilon$ function to include updating a
causal state:
\begin{align*}
s_j &= \epsilon(s_k,y_{k:j}) \\
  & \equiv \{y_{0:l}'|\exists \, y_{0:k} \ni s_k=\epsilon(y_{0:k}) \text{ and } y_{0:l}'\sim y_{0:j}\}
  ~.
\end{align*}
This is the set of histories $y_{0:l}'$ that predict the same future as a
history $y_{0:j}$ which leads to causal state $s_k$ via the initial sequence
$y_{0:k}$ and then follows with the sequence $y_{k:j}$.

Unifilarity is key to deducing several useful properties of the HMM
$\theta^{(y)}_{s \rightarrow s'}$, which we will refer to as a
\emph{nonstationary} \eM.
First, unifilarity implies that for any causal state $s$ followed
by a symbol $y$, there is a unique next state $s'=\epsilon(s,y)$, meaning that
the symbol-labeled transition matrix can be written:
\begin{align*}
\theta^{(y)}_{s \rightarrow s'}
  = \theta^{(y)}_{s \rightarrow \epsilon(s,y)} \delta_{s',\epsilon(s,y)}
  ~.
\end{align*}
Moreover, the \eM's form is uniquely determined from the semi-infinite process:
\begin{align*}
\theta^{(y)}_{s \rightarrow s'}=\Pr(Y^\theta_j=y|S^\theta_j=s)\delta_{s',\epsilon(s,y)}
  ~,
\end{align*}
where the conditional probability is determined from the process:
\begin{align*}
\Pr(Y^\theta_j=y|S^\theta_j=\epsilon(y_{0:i}))
  = \Pr(Y^\theta_j=y|Y^\theta_{0:j}=y_{0:j})
  ~.
\end{align*}
Once constructed, the \eM\ allows us to reconstruct word probabilities via the
simple product:
\begin{align*}
\Pr(Y^\theta_{0:L}=y_{0:L})
  = \prod_{j=0}^{L-1}
  \theta^{(y_j)}_{\epsilon(s^*,y_{0:j}) \rightarrow\epsilon(s^*,y_{0:j+1})}
  ~,
\end{align*}
where $y_{0:0}$ denotes the null word, taking a causal state to itself under
the causal update $\epsilon(s,y_{0:0})=s$.

Allowing for arbitrarily-many causal states, our class of models (nonstationary
\eMs) is so general that it can represent \emph{any} semi-infinite process and,
thus, any distribution over sequences $\mathcal{Y}^L$. One concludes that
computational mechanics provides an ideal class of generative models to fit to
data $y_{0:L}$. \emph{Bayesian structural inference} implements just this
\cite{Stre13a}.

In these ways, computational mechanics already had solved (and several decades
prior) the unsupervised learning challenge recently posed by Ref. \cite{Wu19a}
to create an ``AI Physicist'': a machine that learns regularities in time
series to make predictions of the future from the past \cite{Crut92c}.

\subsubsection{Input-Output Machines}

\begin{figure}[tbp]
\centering
\includegraphics[width=\columnwidth]{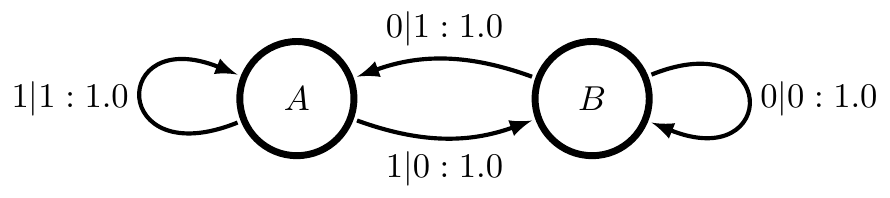}
\caption{The delay channel \eT: The last input symbol is stored in its memory
	(states). If the last
	symbol was $1$, then the corresponding transitions, labeled $y'|1:1.0$,
	update the hidden state to $A$. Then, all outputs from $A$ are symbol $1$.
	Similarly, input $0$ leads to state $B$, whose corresponding outputs are
	all $0$. In this way, the delay channel outputs the previous input
	symbol.
	}
\label{fig:DelayChannel}
\end{figure}

This way one constructs a predictive HMM that generates a desired semi-infinite
process $\Pr(Y^\theta_{0:L}=y_{0:L})$. The generalization to \emph{\eTs} allows for an \emph{input} as well as an output process---a transformation between processes \cite{Barn13a}. The transducer at the $i$th time step is
described by transitions among the hidden states $X_i \rightarrow X_{i+1}$,
that are conditioned on the input $Y_i$, emitting the output $Y_i'$:
\begin{align*}
M_{x \rightarrow x'}^{(y'|y)}=\Pr(Y_i'=y',X_{i+1}=x'|Y_i=y,X_i=x)
  ~.
\end{align*}
\ET  \,\,state-transition diagrams label the edges of transitions between
hidden states $y'|y:M_{x \rightarrow x'}^{(y'|y)}$. As Fig.
\ref{fig:DelayChannel} shows this is to be read as the probability $M_{x
\rightarrow x'}^{(y'|y)}$ of output $y'$ and next hidden state $x'$ given input
$y$ and current hidden state $x$.

These devices are memoryful channels, with their memory encoded in the hidden
states $X_i$. They implement a wide variety of functional operations. Figure
\ref{fig:DelayChannel} shows the \emph{delay channel}. With sufficient memory, though,
an \eT\ can implement a universal Turing machine \cite{Broo89a}. Moreover, if
the input and output alphabets are the same, then they represent the form of a
physical information ratchet, which have energetic requirements that arise from
the thermodynamics of their operation \cite{Mand012a,Boyd15a}. Since these
physically-implementable information processors are so general in their ability
to compute, they represent a very broad class of physical agents. As such, we
use the framework of information ratchets to explore the functionality of
agents that process information as a fuel.

\section{Proof of Zero Entropy Production of Trajectories}
\label{app:zero_ent}

Perfectly-efficient agents dissipate zero work and generate zero entropy
$\langle \Sigma \rangle=0$. The Crooks fluctuation theorem and other detailed
fluctuation theorems say that entropy production is proportional to the
log-ratio of probabilities \cite{Croo99a, Jarz00a}:
\begin{align*}
\Sigma_{|z_{\tau:\tau'}}
  = \kB \ln \frac{\rho_F(\Sigma_{|z_{\tau:\tau'}})}{\rho_R(-\Sigma_{|z_{\tau:\tau'}})}
  ~,
\end{align*}
where $\rho_F(\Sigma_{|z_{\tau:\tau'}})$ is the probability of the entropy
production under the protocol that controls the ratchet and
$\rho_R(-\Sigma_{|z_{\tau:\tau'}})$ is the probability of minus that same
entropy production if the control protocol is reversed. Thus, the average
entropy production is proportional to the relative entropy between these two
distributions \cite{Cove06a}:
\begin{align*}
\langle \Sigma \rangle
  & = \kB \sum_{z_{\tau:\tau'}} \rho_F(\Sigma_{|z_{\tau:\tau'}}) \ln
  \frac{\rho_F(\Sigma_{|z_{\tau:\tau'}})}{\rho_R(-\Sigma_{|z_{\tau:\tau'}})} \\
  & \equiv \kB D_{KL} (\rho_F(\Sigma_{|z_{\tau:\tau'}})
  ||\rho_R(-\Sigma_{|z_{\tau:\tau'}}))
  ~.
\end{align*}
If the control is thermodynamically efficient, this relative entropy vanishes \cite{Jarz11a}, implying the necessary and sufficient condition that $\rho_F(\Sigma)=\rho_R(-\Sigma)$. This then implies that all paths produce
zero entropy:
\begin{align*}
\Sigma_{|z_{\tau:\tau'}}=0.
  ~
\end{align*}
Entropy fluctuations vanish as the entropy production goes to zero.

\section{Thermodynamically Efficient Markov Channels}
\label{app:Thermodynamically Efficient Markov Channels}

Given a physical system $\mathcal{Z}=\{z\}$, a computation on its states is
given by a Markov channel $M_{z \rightarrow z'}=\Pr(Z_{\tau'}=z'|Z_{\tau}=z)$
and an input distribution $\Pr(Z_\tau=z)$. The following  describes a
quasistatic thermodynamic control that implements this computation efficiently
if the input distribution matches the estimated distribution $\Pr(Z^\theta=z)$.
This means that the work production is equal to the change in pointwise
nonequilibrium free energy:
\begin{align}
\left\langle
W^\theta_{|z_{\tau},z_{\tau'}}
\right\rangle
  & =\phi(z_{\tau'},\tau')- \phi(z_\tau,\tau) \\
  & = \Delta E_\mathcal{Z}
  + \kB T \ln
  \frac{\Pr(Z^\theta_{\tau'}=z_{\tau'})}{\Pr(Z^\theta_{\tau}=z_{\tau})}
  ~. \nonumber
\label{eq:EfficientWork3}
\end{align}
Note that, while $\Pr(Z^\theta_\tau=z)$ is the input distribution for which the
computation is efficient, it is possible that other input distributions
$\Pr(Z_\tau=z)$ yield zero entropy production as well. They are only required to minimize
$D_{KL}(Z_\tau||Z^\theta_\tau)-D_{KL}(Z_{\tau'}||Z^\theta_{\tau'})=0$ \cite{Kolc17a}.

The physical setting that we take for thermodynamic control is overdamped
Brownian motion with a controllable energy landscape. This is described by
detailed-balanced rate equations. However, if our physical state-space is
limited to $\mathcal{Z}$, then not all channels can be implemented with
continuous-time rate equations \cite{Ray20b}. Fortunately, this can be
circumvented by additional ancillary or hidden states \cite{Owen19a,Ray20b}.
And so, to implement any possible channel, we add an ancillary copy of our
original system $\mathcal{Z'}$, such that our entire physical system is
$\mathcal{Z}_\text{total}=\mathcal{Z} \times \mathcal{Z}'$.

Prescriptions have been given that efficiently implement any computation,
specified by a Markov channel $M_{z_\tau \rightarrow
z_{\tau'}}=\Pr(Z_{\tau'}=z_{\tau'}|Z_\tau=z_\tau)$, using quasistatic
manipulation of the $\mathcal{Z}$'s energy levels and an ancillary copy
$\mathcal{Z}'$ \cite{Garn15,Boyd17a}. However, these did not determine the work
production for individual computational maps $z_\tau \rightarrow
z_\tau'$ during the computation interval $(\tau, \tau')$.

The following implements an analogous form of quasistatic computation that
allows us to easily calculate the energy associated with implementing the
computation $M_{z_\tau \rightarrow z_{\tau'}}$, assuming the subsystem
$\mathcal{Z}$ started in $z_{\tau}$ and ends in $z_{\tau'}$. Due to detailed
balance, the rate equation dynamics over the computational system and its
ancillary copy $\mathcal{Z}_\text{total}=\mathcal{Z} \times \mathcal{Z}'$ are
partially specified by the energy $E(z,z',t)$ of system state $z$ and ancillary
state $z'$ at time $t$. This also uniquely specifies the equilibrium
distribution:
\begin{align*}
\Pr(Z^\text{eq}_t=z,Z^{'\text{eq}}_t=z')
  = \frac{e^{-E(z,z',t)/ \kB T}}{\sum_{z,z'}e^{-E(z,z',t)/ \kB T}}
  ~.
\end{align*}
The normalization constant $\sum_{z,z'}e^{-E(z,z',t)/ \kB T}$ is the partition
function that determines the equilibrium free energy:
\begin{align*}
F^\text{eq}(t)
  =- \kB T \ln \left( \sum_{z,z'}e^{-E(z,z',t)/\kB T} \right)
  ~.
\end{align*}
The equilibrium free energy adds to the system energy. It is constant over the
states:
\begin{align*}
E(z,z',t)
  = F^\text{eq}(t)- \kB T \ln \Pr(Z^\text{eq}_t=z,Z^{'\text{eq}}_t=z')
  ~.
\end{align*}

\begin{figure}[tbp]
\centering
\includegraphics[width=\columnwidth]{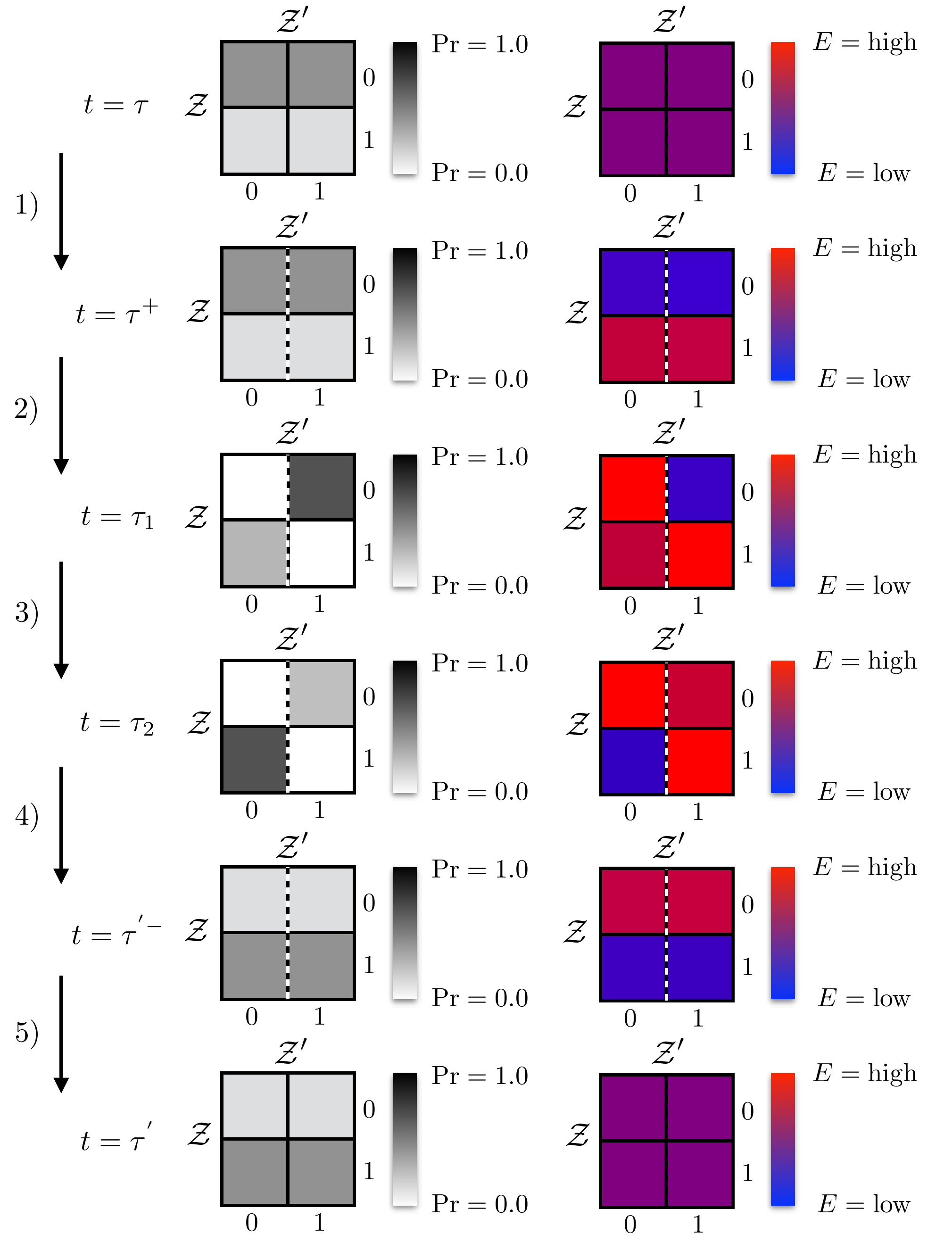}
\caption{Quasistatic agent implementing the Markov chain
	$M_{z_\tau \rightarrow z_\tau'}$ in the system $\mathcal{Z}$ over the time
	interval $[\tau,\tau']$ using ancillary copy $\mathcal{Z}'$ in five steps:
	Epoch 1: Energy landscape is instantaneously brought into equilibrium with
	the distribution over the joint system. Epoch 2: Probability flows in the
	ancillary system $\mathcal{Z}'$ as the energy landscape quasistatically
	changes to make the conditional probability distribution in $\mathcal{Z}'$
	reflect the Markov channel $\Pr(Z'_{\tau_1}=z'|Z_{\tau_1}=z)=M_{z
	\rightarrow z'}$. Epoch 3: Systems $\mathcal{Z}$ and $\mathcal{Z}'$ are
	swapped. Epoch 4: Ancillary system quasistatically reset to the uniform
	distribution. Epoch 5: Energy landscape instantaneously reset to uniform.
	}
\label{fig:QuasistaticAgent}
\end{figure}

We leverage the relationship between energy and equilibrium probability to
design a protocol that achieves the work production given by Eq.
(\ref{eq:EfficientWork3}) for a Markov channel $M$. The estimated distribution
over the whole space assumes that the initial distribution of the ancillary
variable is uncorrelated and uniformly distributed:
\begin{align*}
\Pr(Z^\theta_\tau,Z'_\tau)=\frac{\Pr(Z^\theta_\tau)}{|\mathcal{Z}|}
  ~.
\end{align*}
Assuming the default energy landscape is constant initially and
finally---$E(z,z',\tau)=E(z,z',\tau')=\xi$---the maximally efficient protocol
over the interval $[\tau,\tau']$ decomposes into five epochs, see Fig.
\ref{fig:QuasistaticAgent}:
\begin{enumerate}
      \setlength{\topsep}{-2pt}
      \setlength{\itemsep}{-4pt}
      \setlength{\parsep}{-2pt}
\item Quench: $[\tau,\tau^+]$,
\item Quasistatically evolve: $(\tau,\tau_1]$,
\item Swap: $(\tau_1,\tau_2]$,
\item Quasistatically evolve: $(\tau_2,\tau')$, and
\item Reset: $[\tau^{'-},\tau']$.
\end{enumerate}
For all protocol epochs, except for Epoch
3 during which the two subsystems are swapped, $\mathcal{Z}$ is held fixed
while the ancillary system $\mathcal{Z}'$ follows the local equilibrium
distribution. Let's detail these in turn.

\paragraph*{1. Quench:}
Instantaneously quench the energy from $E(z,z',\tau)=\xi$ to $E(z,z',\tau^+)=
\kB T \ln (|\mathcal{Z}|/\Pr(Z_\tau=z))$ over the infinitesimal time interval
$[\tau, \tau^+]$ such that, if the distribution was as we expect, it would be
in equilibrium $\Pr(Z^\text{eq}_\tau,Z^{'\text{eq}}_\tau)=\Pr(Z^\theta_\tau)/|\mathcal{Z}|$.

If the system started in $z_\tau$, then the associated work produced is
opposite the energy change:
\begin{align*}
\left\langle
W^{\theta,1}_{|z_\tau,z_{\tau'}}
\right\rangle
  & = E(z_\tau,z',\tau)-E(z_\tau,z',\tau^+)  \nonumber \\
  & = \xi + \kB T \ln \frac{\Pr(Z^\theta_\tau=z_\tau)}{|\mathcal{Z}|}
  ~.
\end{align*}
$\left\langle W^{\theta,1}_{|z_\tau,z_{\tau'}}
\right\rangle$ denotes that the work is produced in Epoch 1, conditioned on
the estimated distributions $Z^\theta_\tau$ and $Z^\theta_{\tau'}$, and initial
and final states $z_\tau$ and $z_{\tau'}$. Note that we also condition on
$Z^\theta_{\tau'}=z_{\tau'}$, since work production in this phase is
unaffected by the computation's end state.

\paragraph*{2. Quasistatically evolve:}
Quasistatically evolve the energy landscape over a third of total time interval
$(\tau, \tau_1]$ such that the joint system remains in equilibrium and the
ancillary system $\mathcal{Z}'$ is determined by the Markov channel $M$ applied
to the system $\mathcal{Z}$:
\begin{align*}
\Pr(Z_{\tau_1}=z,Z'_{\tau_1}=z')&=\Pr(Z_\tau=z)M_{z \rightarrow z'} \\
  E(z,z', \tau_1)& =- \kB T \ln \Pr(Z^\theta_\tau=z)M_{z \rightarrow z'}
  ~.
\end{align*}
Also, hold the energy barriers between states in $\mathcal{Z}$ high, preventing
probability flow between states and preserving the distribution $\Pr(Z_t)=\Pr(Z_\tau)$ for all $t \in ( \tau, \tau_1]$.

Given that the system started in $Z_\tau=z_\tau$, the work production during
this epoch corresponds to the average change in energy:
\begin{align*}
& \left\langle
  W^{\theta,2}_{|z_\tau,z_{\tau'}}
  \right\rangle \\
  & = - \sum_{z,z'}\int_{\tau^+}^{\tau_1} dt
  \Pr(Z'_t=z',Z_t=z|Z_\tau=z_\tau)  \partial_{t}E(z,z',t)
  ~.
\end{align*}
As the system $\mathcal{Z}$ remains in $z_\tau$ over the interval:
\begin{align*}
\Pr(Z'_t=z',Z_t=z|Z_\tau=z_\tau)=\Pr(Z'_t=z'|Z_t=z)\delta_{z,z_\tau}
\end{align*}
and the work production simplifies to:
\begin{align*}
& \left\langle
W^{\theta,2}_{|z_\tau,z_{\tau'}}
\right\rangle \\
  & \quad = - \sum_{z'}\int_{\tau^+}^{\tau_1} dt
  \Pr(Z'_t=z'|Z_t=z_\tau)  \partial_{t}E(z_\tau,z',t)
  ~.
\end{align*}
We can express the energy in terms of the estimated equilibrium probability
distribution:
\begin{align*}
E(z_\tau,z',t)= - \kB T \ln \Pr(Z'_t=z'|Z_t=z_\tau) \Pr(Z^\theta_t=z_\tau)
  ~.
\end{align*}
And, since the distribution over the system $\mathcal{Z}$ is fixed during this
interval:
\begin{align*}
\Pr(Z'_t=z'|Z_t=z_\tau) & \Pr(Z^\theta_t=z_\tau) \\
  & =\Pr(Z'_t=z'|Z_t=z_\tau)\Pr(Z^\theta_\tau=z_\tau)
  ~.
\end{align*}

Plugging these into the expression for the work production, we find that the
evolution happens without energy exchange:
\begin{align*}
& \left\langle
W^{\theta,2}_{|z_\tau,z_{\tau'}}
\right\rangle \\
   & =- \kB T\int_{\tau^+}^{\tau_1} dt
   \sum_{z'}\Pr(Z'_t=z'|Z_t=z_\tau) \\
   & \qquad \qquad \times \partial_{t} \ln \Pr(Z'_t=z'|Z_t=z_\tau)\Pr(Z^\theta_\tau=z_\tau) \\
   & =- \kB T\int_{\tau^+}^{\tau_1} dt
   \sum_{z'} \Pr(Z'_t=z'|Z_t=z_\tau) \\
   & \qquad \qquad \times \frac{\Pr(Z^\theta_\tau=z_\tau)\partial_{t} \Pr(Z'_t=z'|Z_t=z_\tau)}{\Pr(Z'_t=z'|Z_t=z_\tau)\Pr(Z^\theta_\tau=z_\tau)} \nonumber \\
   & = - \kB T\int_{\tau^+}^{\tau_1} dt\,\sum_{z'}\partial_{t} \Pr(Z'_t=z'|Z_t=z_\tau) \nonumber \\
   & = - \kB T\int_{\tau^+}^{\tau_1} dt\,\partial_{t} \sum_{z'} \Pr(Z'_t=z'|Z_t=z_\tau) \nonumber \\
   & = - \kB T\int_{\tau^+}^{\tau_1} dt\,\partial_{t} 1 \\
   & = 0
   ~.
\end{align*}
After this state, at time $\tau_1$ the resulting joint distribution over the
ancillary and primary system matches the desired computation:
\begin{align}
\Pr(Z_{\tau_1}=z,Z_{\tau_1}'=z')=\Pr(Z_\tau=z)M_{z \rightarrow z'}
  ~.
\end{align}

\paragraph*{3. Swap:}
Over time interval $(\tau_1,\tau_2]$, efficiently swap the two systems
$\mathcal{Z} \leftrightarrow \mathcal{Z}'$, such that
$\Pr(Z_{\tau_1}=z,Z'_{\tau_1}=z')=\Pr(Z_{\tau_2}=z',Z'_{\tau_2}=z)$ and
$E(z,z',\tau_1)=E(z',z,\tau_2)$. This operation requires zero work as well,
as it is reversible, regardless of where the system starts or ends:
\begin{align*}
\left\langle W^{\theta,3}_{|z_\tau,z_{\tau'}}
\right\rangle
  = 0
  ~.
\end{align*}
Such an efficient swap operation has been demonstrated in finite time \cite{Ray20b}.  The resulting joint distribution over the ancillary and primary system matches
a flip of the desired computation:
\begin{align*}
\Pr(Z_{\tau_2}=z,Z_{\tau_2}'=z')=\Pr(Z_\tau=z')M_{z' \rightarrow z}
  ~.
\end{align*}

\paragraph*{4. Quasistatically evolve:}
Over time interval $(\tau_2, \tau')$, quasistatically evolve the energy
landscape from:
\begin{align*}
E(z,z', \tau_2)& =- \kB T \ln \Pr(Z^\theta_\tau=z')M_{z' \rightarrow z}
\end{align*}
to
\begin{align*}
E(z,z', \tau^{'-})& =- \kB T \ln
 \frac{\sum_{z''}\Pr(Z^\theta_\tau=z'')M_{z'' \rightarrow z}}{|\mathcal{Z}|} \\
  & =- \kB T \ln \frac{\Pr(Z^\theta_{\tau'}=z)}{|\mathcal{Z}|}
  ~.
\end{align*}
We keep the primary system $\mathcal{Z}$ fixed as in Epoch 2. And, as in Epoch
2, there is zero work production:
\begin{align*}
\left\langle
W^{\theta,4}_{|z_\tau,z_{\tau'}}
\right\rangle=0
  ~.
\end{align*}
The result is that the primary system is in the desired final distribution:
\begin{align*}
\Pr(Z_{\tau'}=z)\equiv \sum_{z'}\Pr(Z_\tau=z')M_{z' \rightarrow z}
  ~,
\end{align*}
having undergone a mapping from its original state at time $\tau$, while the
ancillary system has returned to an uncorrelated uniform distribution.

\paragraph*{5. Reset:}
Finally, over time interval $[\tau^{'-},\tau']$ instantaneously change the
energy to the default flat landscape $E(z,z',\tau')= \xi$. Given that the
system ends in state $z_{\tau'}$, the associated work production is:
\begin{align*}
\left\langle W^{\theta,5}_{|z_\tau,z_{\tau'}}
\right\rangle
  & = E(z_\tau',z',\tau^{'-})-E(z_\tau',z',\tau') \\
  & = -\xi- \kB T \ln \frac{\Pr(Z^\theta_{\tau'}=z_{\tau'})}{|\mathcal{Z}|}
  ~.
\end{align*}

The net work production given the initial state $z_\tau$ and final state
$z_{\tau'}$ is then:
\begin{align*}
\left\langle
W^\theta_{|z_\tau,z_{\tau'}}
\right\rangle
  & = \sum_{\text{i}=1}^5
  \left\langle W^{\theta,i}_{|z_\tau,z_{\tau'}}
  \right\rangle \\
  & = \kB T \ln
  \frac{\Pr(Z^\theta_{\tau}=z_{\tau})}{\Pr(Z^\theta_{\tau'}=z_{\tau'})}
  .
\end{align*}

Thus, when we average over all possible inputs and outputs and the estimated an
actual distributions are the same $Z^\theta_\tau=Z_\tau$, we see that this
protocol achieves the thermodynamic Landauer's bound:
\begin{align*}
\left\langle W^\theta \right\rangle
  &  \! =  \! \sum_{z_\tau,z_{\tau'}}
  \Pr(Z_\tau \! = \! z_\tau, Z_{\tau'} \! = \! z_{\tau'})
  \left\langle W^\theta_{|z_\tau,z_{\tau'}} \right\rangle \\
  &=  \kB T \ln 2 (H[Z_{\tau'}]-H[Z_{\tau}])
  ~.
\end{align*}
One concludes that this is a thermodynamically-efficient method for computing
any Markov channel.

\section{Designing Agents}
\label{app:Designing Agents}

To design a predictive thermodynamic agent, its hidden states must match the
states of the \eM\ at every time step $X_i=S^\theta_i$. To do this, the agent
states and causal states occupy the same space $\mathcal{X}=\mathcal{S}$, and
the transitions within the agent $M$ are directly drawn from causal equivalence
relation:
\begin{align*}
M_{xy \rightarrow x'y'}= \frac{1}{|\mathcal{Y}|} \times
\begin{cases}
& \delta_{x',\epsilon(x,y)}
  \text{ if } \sum_{x'}\theta^{(y)}_{x \rightarrow x'} \neq 0, \\
  & \delta_{x',x} \text{ otherwise}
  ~.
\end{cases}
\end{align*}
The factor $1/|\mathcal{Y}|$ maximizes work production by mapping to uniform
outputs.

The second term on the right is the probability of the next agent state given
the current input and current hidden state $\Pr(X_{i+1}=x'|Y_i=y,X_i=x)$. The
top case $\delta_{x',\epsilon(x,y)}$ gives the probability that the next
\emph{causal} state is $S^\theta_{i+1}=x'$ given that the current causal state
is $S^\theta_i=x$ and output of the $\epsilon$-machine is $Y_i=y$. This is
contingent on the probability of seeing $y$ given causal state $x$ being
nonzero. If it is, then the transitions among the agent's hidden states match
the transitions of the \eM's causal states.

In this way, if $y_{0:L}$ is a sequence that could be produced by the \eM, we
have designed the agent to stay synchronized to the causal state of the input
$X_i=S^\theta_i$, so that the ratchet is predictive of the process
$\Pr(Y^\theta_{0:\infty})$ and produces maximal work by fully randomizing the
outputs:
\begin{align*}
\Pr(Y'_i=y'_i|\cdot)=\frac{1}{|\mathcal{Y}|}
  ~.
\end{align*}

It can be the case that the \eM\ cannot produce $y$ from the causal state $x$.
This corresponds to a disallowed transition of our model $\theta^{(y)}_{x
\rightarrow x'}=0$. In these cases, we arbitrarily choose the next state to be
the same $\delta_{x,x'}$. There are many possible choices, though---such
as resetting to the start state $s*$. However, it is physically irrelevant,
since these transitions correspond zero estimated probability and, thus, to
infinite work dissipation, drowning out all other details of the model.
However, this particular choice for when $y$ cannot be generated from the
causal state $x$ preserves unifilarity and allows the agent to wait in its
current state until it receives an input that it can accept.

Modulo disallowed, infinitely-dissipative transitions, we now have a direct
mapping between our estimated input process $\Pr(Y^\theta_{0:\infty})$ and its
\eM\ $\theta$ to the logical architecture $M$ of a maximum work-producing agent.

As yet, this does not fully specify agent behavior, since it leaves out the
estimated input distribution $\Pr(Y^\theta_i=y,X^\theta_i=x)$. This
distribution must match the actual distribution $\Pr(Y_i=y,X_i=x)$ for the
agent to be locally efficient, not accounting for temporal correlations.
Fortunately, since agent states are designed to match the \eM's causal states,
we know that the agent state distribution matches the causal-state distribution
and inputs:
\begin{align*}
\Pr(Y^\theta_i=y_i,X^\theta_i=s_i)& =\Pr(Y^\theta_i=y_i,S^\theta_i=s_i)
  ~,
\end{align*}
if the ratchet is driven by the estimated input. The joint distribution over
causal states and inputs is also determined by the \eM, since the construction
assumes starting in the state $s_0=s^*$. To start, note that the joint
probability trajectory distribution is given by:
\begin{align*}
\Pr(Y^\theta_{0:i+1} & =y_{0:i+1},S^\theta_{0:i+2}=s_{0:i+2}) \\
  & = \delta_{s_0,s^*} \prod_{j=0}^{i} \theta^{(y_j)}_{s_i \rightarrow s_{i+1}}
  ~.
\end{align*}
Summing over the variables besides $Y^\theta_i$ and $S^\theta_i$, we obtain an
expression for the estimated agent distribution in terms of just the \eM's HMM:
\begin{align*}
\Pr(Y^\theta_i=y_i,X^\theta_i=s_i)
  = \sum_{y_{0:i},s_{0:i},s_{i+1}} \delta_{s_0,s^*}
  \prod_{j=0}^{i} \theta^{(y_j)}_{s_j \rightarrow s_{j+1}}
  ~.
\end{align*}
Thus, an agent $\{M,\Pr(X^\theta_i,Y^\theta_i)\}$ designed to be \emph{globally
efficient} for the estimated input process $\Pr(Y^\theta_{0:L})$ can be derived
from the estimated input process through its \eM\ $\theta^{(y)}_{s \rightarrow
s'}$.

\section{Work Production of Optimal Transducers}
\label{app:Work Production of Optimal Transducers}

The work production of an arbitrary transducer $M$ driven by an input $y_{0:L}$
can be difficult to calculate, as shown in Eq.  (\ref{eq:TransducerWork}).
However, when the transducer is designed to harness an input process with \eM\
$T$, such that:
\begin{align*}
M_{xy \rightarrow x'y'}= \frac{1}{|\mathcal{Y}|} \times
\begin{cases}
&\delta_{x',\epsilon(x,y)}
  \text{ if } \sum_{x'}\theta^{(y)}_{x \rightarrow x'} \neq 0, \\
  & \delta_{x',x} \text{ else},
\end{cases}
\end{align*}
the work production simplifies. To see this, we express Eq.
(\ref{eq:TransducerWork}) in terms of the estimated  distribution
$\Pr(Y^\theta_{0:L})$, ratchet $M$, and input $y_{0:L}$, assuming that the word
$y_{0:L}$ can be produced from every initial hidden state with nonzero
probability $\Pr(Y^\theta_{0:L}=y_{0:L}|S_0=s_0) \Pr(X_0=s_0)\neq 0$, which
guarantees that $ \sum_{x_{i+1}}\theta^{(y_i)}_{x_i \rightarrow x_{i+1}} \neq
0$. If this constraint is not satisfied, the agent will dissipate infinite
work, as it implies $\Pr(X^\theta_i=x_i,Y^\theta_i=x_i)=0$ for some $i$. Thus,
we use the expression $M_{xy \rightarrow
x'y'}=\delta_{x',\epsilon(x,y)}/|\mathcal{Y}|$ in the work production:
\begin{widetext}
\begin{align*}
& \left\langle W^\theta_{|y_{0:L}} \right\rangle \\
  & \quad = \kB T \sum_{x_{0:L+1},y_{0:L}'} \Pr(X_0=x_0)
  \prod_{j=0}^{L-1}M_{x_j,y_j \rightarrow x_{j+1},y_j'}
  \ln \prod_{i=0}^{L-1}
  \frac{\Pr(X^\theta_i=x_i,Y^\theta_i=y_i)}{\Pr(X^\theta_{i+1}=x_{i+1},Y^{'\theta}_i=y'_i)} \\
  & \quad = \kB T \sum_{x_{0:L+1},y_{0:L}'} \Pr(X_0=x_0)
  \prod_{j=0}^{L-1} \frac{\delta_{x_{j+1},\epsilon(x_j,y_j)}}{|\mathcal{Y}|}
  \ln \prod_{i=0}^{L-1}
  \frac{\Pr(X^\theta_i=x_i,Y^\theta_i=y_i)}{\Pr(X^\theta_{i+1}=x_{i+1})/|\mathcal{Y}|} \\
  & \quad = \kB T \ln |\mathcal{Y}|
  + \kB T \sum_{x_{0:L+1}} \Pr(X_0=x_0)
  \prod_{j=0}^{L-1} \delta_{x_{j+1},\epsilon(x_j,y_j)}
  \left( \ln \prod_{i=0}^{L-1} \Pr(Y^\theta_i=y_i|X^\theta_i=x_i)
  + \ln \prod_{i=0}^{L-1} \frac{\Pr(X^\theta_i=x_i)}{\Pr(X^\theta_{i+1}=x_{i+1})} \right)
  ~.
\end{align*}

Note that $\prod_{j=0}^{L-1} \delta_{x_{j+1},\epsilon(x_j,y_j)}$ vanishes
unless each element of the hidden state trajectory $x_i$ corresponds to the
resulting state of the \eM\ when the initial state $x_0$ is driven by the
first $i$ inputs $y_{0:i}$, which is $\epsilon(x_0,y_{0:i})$. The fact that
the agent is driven into a unique state is guaranteed by the \eM's unifilarity.
Thus, we rewrite:
\begin{align*}
\prod_{j=0}^{L-1} \delta_{x_{j+1},\epsilon(x_j,y_j)}
 = \prod_{j=1}^{L} \delta_{x_j,\epsilon(x_0,y_{0:j})}
  ~.
\end{align*}
This engenders a simplification of the work production:
\begin{align}
\left\langle W^\theta_{|y_{0:L}} \right\rangle
  & = \kB T \ln |\mathcal{Y}|
  + \kB T \sum_{x_{0:L+1}} \Pr(X_0=x_0)
  \prod_{j=1}^{L} \delta_{x_j,\epsilon(x_0,y_{0:j})}
  \ln \prod_{i=0}^{L-1}
  \frac{\Pr(Y^\theta_i=y_i|X^\theta_i=x_i)\Pr(X^\theta_i=x_i)}
  {\Pr(X^\theta_{i+1}=x_{i+1})} \nonumber \\
  & = \kB T \ln |\mathcal{Y}|
  + \kB T \sum_{x_0}\Pr(X_0=x_0)
  \ln  \prod_{i=0}^{L-1}
  \Pr(Y^\theta_i=y_i|X^\theta_i=\epsilon(x_0,y_{0:i})) \nonumber \\
  & + \kB T\sum_{x_0} \Pr(X_0=x_0)
  \ln  \frac{\Pr(X^\theta_0=x_0)}{\Pr(X^\theta_{L}=\epsilon(x_0,y_{0:L}))}
\label{eq:PerformanceMeasure}
  ~,
\end{align}
\end{widetext}
where $y_{0:0}$ denotes the null input, which leaves the state fixed under the $\epsilon$-map $\epsilon(x_0,y_{0:0})=x_0$.

This brings us to an easily calculable work production, especially when the
system is initialized in the start state $s^*$. Recognizing that if we
initiate the \eM\ in its start state $s^*$, such that
$\Pr(X_0=x_0)=\delta_{x_0,s^*}$, then $X_0=S_0$ is predictive. By extension,
every following agent state is predictive and equivalent to the causal state
$X_i=S_i$ yielding:
\begin{align*}
 \Pr(Y_i=y_i|X_i=\epsilon(s^*,y_{0:i}))
 & = \Pr(Y_i=y_i|S_i=\epsilon(s^*,y_{0:i}))
 \\
 & = \Pr(Y_i=y_i|Y_{0:i}=y_{0:i})
 ~.
\end{align*}
Thus, the work production simplifies a sum of terms that includes the
log-likelihood of the finite input:
\begin{align*}
& \left\langle W^\theta_{|y_{0:L}} \right\rangle  \\
  & = \kB T \sum_{x_0} \delta_{x_0,s^*}
  \ln \prod_{i=0}^{L-1} \Pr(Y^\theta_i=y_i|Y^\theta_{0:i}=y_{0:i}) \\
  & \quad + \kB T L \ln |\mathcal{Y}|+ \kB T \sum_{x_0} \delta_{x_0,x^*}
  \ln \frac{\delta_{x_0,s^*}}{\Pr( X^\theta_L=\epsilon(x_0,y_{0:L}))} \\
  & = \kB T (\ln \Pr(Y^\theta_{0:L}=y_{0:L})+  L \ln |\mathcal{Y}| \\
  & \quad - \ln \Pr( X^\theta_L=\epsilon(s^*,y_{0:L})))
  ~.
\end{align*}

The first $\log(\cdot)$ in the last line is the log-likelihood of the model
generating $y_{0:L}$---a common performance measure for machine learning
algorithms. If an input has zero probability this leads to $-\infty$ work
production, and all other features are drowned out by the log-likelihood term.
Thus, the additional terms that come into play when the input probability
vanishes become physically irrelevant: the agent is characterized by the \eM.
From a machine learning perspective, the model is also characterized by the
\eM\ $\theta^{(y)}_{s \rightarrow s'}$ for the process
$\Pr(Y^\theta_{0:\infty})$. The additional term $\kB T L \ln |\mathcal{Y}|$ is
the work production that comes from exhausting fully randomized outputs and
does not change depending on the underlying model.

The final term $-\kB T \ln \Pr( X^\theta_L=\epsilon(s^*,y_{0:L})))$ does
directly depend on the model. $\Pr(X^\theta_L=x)$ is the distribution over
agent states $\mathcal{X}$ at time $L\tau$ if the agent is driven by the
estimated input distribution $Y^\theta_{0:L}$. This component of the work
production is larger, on average, for agents with high state uncertainty, since
this leads, on-average, to smaller values of $\Pr( X^\theta_L)$. This
contribution to the work production comes from the state space expanding from
the start state $s^*$ to the larger (recurrent) subset of agent states, and so
it provides additional work. This indicates that we are neglecting the cost of
resetting to the start state while harnessing the energetic benefit of starting
in it.

If the machine is designed to efficiently harness inputs again after it
operates on one string, it must be reset to the start state $s^*$. This can
be implemented with an efficient channel that anticipates the input distribution
$\Pr(X^\theta_L=x)$, outputs the distribution
$\Pr(X^\theta_{L+1}=x)=\delta_{x,s^*}$, and so costs:
\begin{align*}
 W^{\theta,\text{reset}}_{|y_{0:L}}
   = \kB T \ln \Pr(X^\theta_L=\epsilon(s^*,y_{0:L}))
  ~.
\end{align*}
Thus, when we add the cost of resetting the agent to the start state at
$X_{L+1}$, the work production is dominated by the log-likelihood:
\begin{align}
\left\langle W^\theta_{|y_{0:L}} \right\rangle
  \! = \! \kB T (\ln \Pr(Y^\theta_{0:L}\! =\! y_{0:L}) \! + \! L \ln |\mathcal{Y}|)
  ~.
\label{eq:EfficientWorkProduction}
\end{align}

\section{Training Simple Agents}
\label{sec:Training Memoryless Agents}

We now outline a case study of thermodynamic learning that is experimentally
implementable using a controllable two-level system. We first introduce a
straightforward method to implement the simplest possible efficient agent.
Second, we show that this physical process achieves the general
maximum-likelihood result arrived at in the main development. Last, we find
the agent selected by thermodynamic learning along with its corresponding
model. As expected, we find that this maximum-work producing agent learns
its environment's predictive features.

\subsection{Efficient Computational Trajectories}

\begin{figure}[tbp]
\centering
\includegraphics[width=.6\columnwidth]{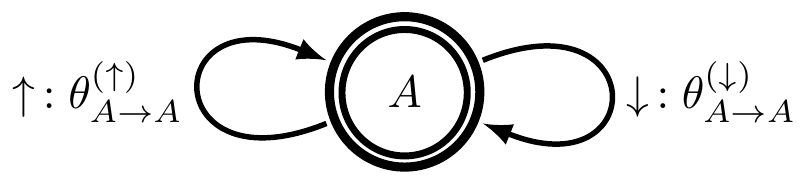}
\caption{Memoryless model of binary data consisting of a single state $A$ and
	the probability of outputting a $\uparrow$ and a $\downarrow$, denoted
	$\theta^{(\uparrow)}_{A \rightarrow A}$ and $\theta^{(\downarrow)}_{A
	\rightarrow A}$, respectively.
	}
\label{fig:Memoryless}
\end{figure}

The simplest possible information ratchets have only a single internal state
$A$ and receive binary data $y_j$ from a series of two-level systems
$\mathcal{Y}_j=\{\uparrow,\downarrow\}$. These agents' internal models
correspond to memoryless \eM s, as shown in Fig. \ref{fig:Memoryless}. The
model's parameters are the probabilities of emitting $\uparrow$ and
$\downarrow$, denoted $\theta^{(\uparrow)}_{A \rightarrow A}$ and
$\theta^{(\downarrow)}_{A \rightarrow A}$, respectively.

Our first step is to design an efficient computation that maps an input distribution
$\Pr(Z_{j\tau})$ to an output distribution $\Pr(Z_{j\tau+\tau'})$ over the
$j$th interaction interval $[j\tau,j\tau+\tau']$. The agent corresponds to the
Hamiltonian evolution $\mathcal{H}_\mathcal{Z}(t)=\mathcal{H}_{\mathcal{X}
\times \mathcal{Y}_j }(t)$ over the joint space of the agent memory and $j$th
input symbol. The resulting energy landscape $E(z,t)$ is entirely specified by
the energy of the two input states $E( A \, \times \uparrow,t)$ and $E(A\,
\times \downarrow, t)$.

\begin{figure*}[tbp]
\centering
\includegraphics[width=1.6\columnwidth]{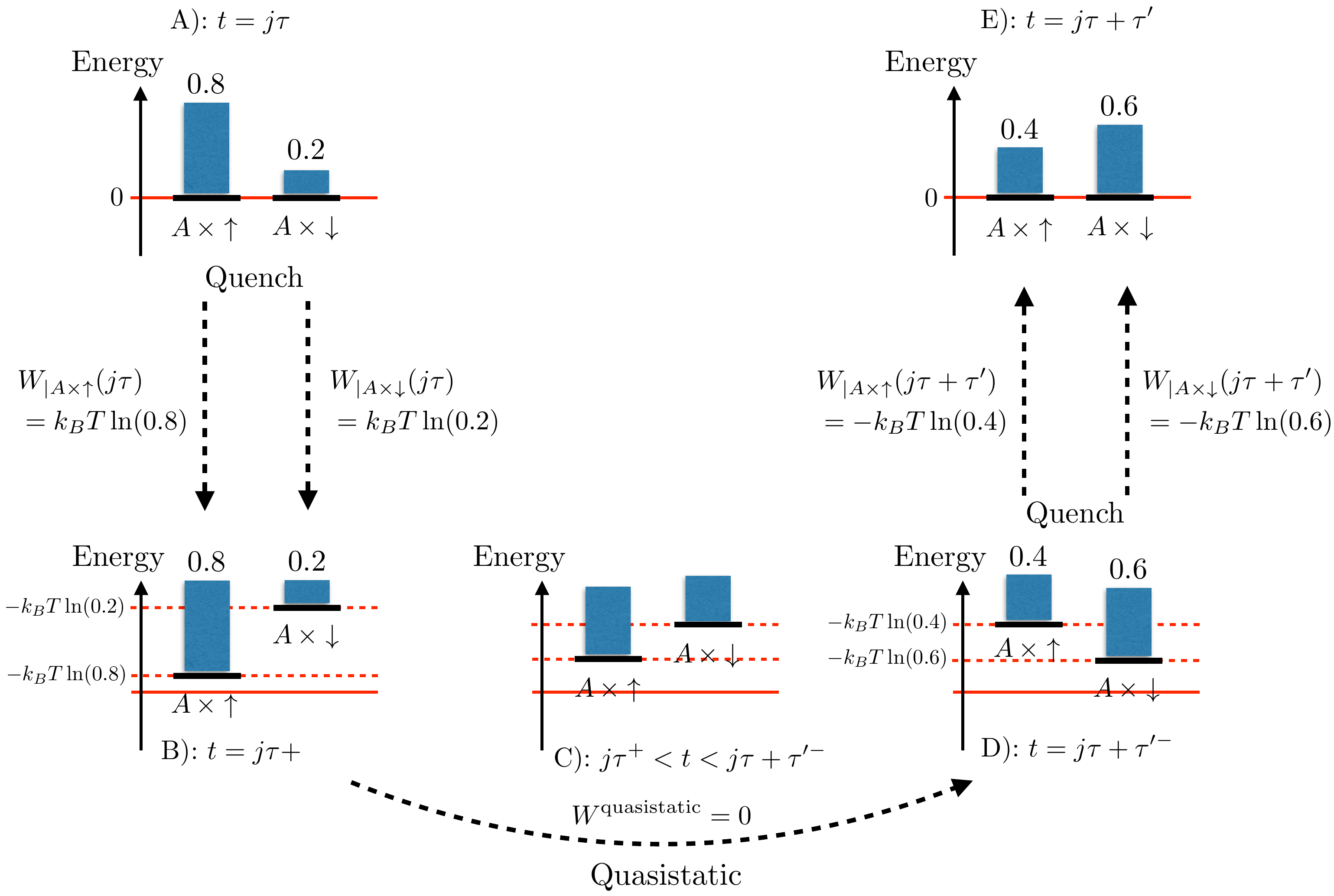}
\caption{Joint two-level system $\mathcal{Z}=\mathcal{X}\times
	\mathcal{Y}_j=\{A\,\times \uparrow,A\,\times \downarrow\}$ undergoing
	perfectly-efficient computation when it receives its estimated input
	through a series of operations. The computation occurs over the time
	interval $t \in (j \tau,j\tau+\tau')$. At panel A) $t=j\tau$ and the system
	has a default flat energy landscape energy $E(z ,j \tau)=E(x\times y ,j
	\tau)=0$. However, it is out of equilibrium, since it is in the
	distribution $\Pr(Z^\theta_{j\tau}=\{A\, \times \uparrow, A\, \times
	\downarrow \})=\{0.8, 0.2 \}$. The first operation is a quench, which
	instantaneously sets the energies be in equilibrium with the initial
	distribution, as shown in panel B). The associated energy change is work.
	Then, a quasistatic operation slowly evolves the system in equilibrium,
	through panel C), to the final desired distribution
	$\Pr(Z^\theta_{j\tau+\tau'}=\{A\, \times \uparrow, A\, \times \downarrow
	\})=\{0.4, 0.6 \}$, shown in panel D). This requires no work. Then, the
	final operation is another quench, in which the energies are reset to the
	default energy landscape $E(z,j\tau+\tau')=0$, leaving the system as shown
	in panel E). Again, the change in energy corresponds to work invested
	through control. The total work production for a particular computational
	mapping $A\times y \rightarrow A\times y'$ is given by the work from the
	initial quench $W_{|A\times y}(j\tau)$ plus the work from the final quench
	$W_{|A\times y'}(j\tau+\tau')$.
	}
\label{fig:WorkEfficient}
\end{figure*}

Appropriately designing this energy landscape allows us to implement the
efficient computation shown in Fig. \ref{fig:WorkEfficient}. The thermodynamic
evolution there instantaneously quenches the energy landscape into equilibrium
with the estimated distribution at the beginning of the interaction interval
$\Pr(Z^\theta_{j\tau})$, then quasistatically evolves the system in equilibrium
to the estimated final distribution $\Pr(Z^\theta_{j\tau+\tau'})$, and,
finally, quenches back to the default energy landscape. In Fig.
\ref{fig:WorkEfficient}, the system undergoes a cycle, starting and ending with
the same flat energy landscape, such that $\Delta E_\mathcal{Z}=0$. This cycle
evolves the distribution over the joint states $A\, \times \uparrow$ and $A\,
\times \downarrow$ from $\Pr(Z^\theta_{j\tau}=\{A\, \times \uparrow, A\, \times
\downarrow \})=\{0.8, 0.2 \}$ to $\Pr(Z^\theta_{j\tau+ \tau'}=\{A\, \times
\uparrow, A\, \times \downarrow \})=\{0.4, 0.6 \}$. Note that this strategy can
be used to evolve between any initial and final distributions.

We control the transformation over time interval $t \in (j\tau,j\tau+\tau')$
such that the time scale of equilibration in the system of interest is much
shorter than the interval length $\tau'$. This slow-moving quasistatic control
means that the states are in equilibrium with the energy landscape over the
interval. In this case, the state distribution becomes the Boltzmann
distribution:
\begin{align*}
\Pr(Z_t=z)=e^{(F^\text{EQ}(t)-E(z,t))/\kB T}
  ~.
\end{align*}
To minimize dissipation for the estimated distribution, the state distribution
must be the estimated distribution $\Pr(Z_t=z)=\Pr(Z^\theta_t=z)$. And so, we
set the two-level-system energies to be in equilibrium with the estimates:
\begin{align*}
E(z,t) = -\kB T \ln \Pr(Z^\theta_t=z)
  ~.
\end{align*}
The resulting process produces zero work:
\begin{align*}
W^\text{quasistatic}
  & = -\int_{z=j\tau}^{j\tau+\tau'} dt\sum_{z}
  \Pr(Z^\theta_t=z) \partial_t E(z,t) \\
  & = 0
\end{align*}
and maps $\Pr(Z^\theta_{j\tau})$ to $\Pr(Z^\theta_{j\tau+\tau'})$ without dissipation.

With the quasistatic transformation producing zero work, the total work
produced from the initial joint state $x\times y$ is exactly opposite the
change in energy during the initial quench:
\begin{align*}
E(x\times y,j\tau) & -E(x\times y,j\tau^+)
  = \kB T \ln \Pr(Z^\theta_{j\tau}=x\times y)
\end{align*}
minus the change in energy of the final joint state $x'\times y'$ during the
final quench:
\begin{align*}
E(x' \times y',j\tau+\tau'^{-})&-E(x'\times y',j\tau+\tau') \\
  & = -\kB T \ln \Pr(Z^\theta_{j\tau+\tau'}=x'\times y')
  ~.
\end{align*}
The two-level system's state is fixed during the instantaneous energy changes.
Thus, if the joint state follows the computational mapping $x\times y
\rightarrow x'\times y'$ the work production is, as expected, directly
connected to the estimated distributions:
\begin{align}
\left\langle W_{|x \times y, x'\times y'} \right\rangle
  = \kB T \ln \frac{\Pr(Z^\theta_{j\tau}=x \times y)}
  {\Pr(Z^\theta_{j\tau+\tau'}=x' \times y')}
  ~.
\label{eq:ExampleWork}
\end{align}
Recall from Sec. \ref{sec:Work Production of Thermodynamic Agents} that the
ratchet system variable $Z^\theta_{j\tau}=X^\theta_j \times Y^\theta_j$ splits
into the random variable $X^\theta_j$ for the $j$th agent memory state and the
$j$th input $Y^\theta_j$. Similarly,
$Z^\theta_{j\tau+\tau'}=X^\theta_{j+1}\times Y^{'\theta}_j$ splits into the
$(j+1)$th agent memory state $X^\theta_{j+1}$ and $j$th output $Y^{'\theta}_j$.
This work production achieves the efficient limit for a model $\theta$ $\left(
\left\langle W_{|x \times y, x'\times y'} \right\rangle=\left\langle
W^\theta_{|x \times y, x'\times y'} \right\rangle \right)$ described in Eq.
(\ref{eq:ModelWork2}).

Appendix \ref{app:Thermodynamically Efficient Markov Channels} generalized the
thermodynamic operation above to \emph{any} computation $M_{z_\tau \rightarrow
z_{\tau'}}$.  While it requires an ancillary copy of the system $\mathcal{Z}$
to execute the conditional dependencies in the computation, it is conceptually
identical in that it uses a sequence of quenching, evolving quasistatically,
and then quenching again.  This appendix extends the strategies outlined in
Refs. \cite{Boyd17a, Garn15} to computational-mapping work calculations.

\subsection{Efficient Information Ratchets}

With the method for efficiently mapping inputs to outputs in hand, we can
design a series of such computations to implement a simple information ratchet
that produces work from a series $y_{0:L}$. As prescribed in Eq.
(\ref{eq:Agents1FromModel}) of Sec. \ref{sec:Designing Agents}, to produce the
most work from estimated model $\theta$, the agent's logical architecture
should randomly map every state to all others:
\begin{align*}
M_{xy \rightarrow x'y'} & = \frac{1}{|\mathcal{Y}_j|} \\
  & = \frac{1}{2}
  ~,
\end{align*}
since there is only one causal state $A$. In conjunction with Eq.
(\ref{eq:ModelFromAgent}), we find that the estimated joint distribution of the
agent and interaction symbol at the start of the interaction is equivalent to
the parameters of the model:
\begin{align*}
\Pr(Z^\theta_{j\tau}=x\times y) & = \Pr(X^\theta_j=x,Y^\theta_j=y) \\
  & = \Pr(Y^\theta_j=y|X^\theta_j=A) \Pr(X^\theta_j=A) \\
  & = \theta^{(y)}_{A \rightarrow A}
  ~,
\end{align*}
where we again used the fact that $A$ is the only causal state. In turn, the
estimated distribution after the interaction is:
\begin{align*}
\Pr(Z^\theta_{j\tau+\tau'}=x'\times y')
  & = \sum_{xy}\Pr(X^\theta_j=x,Y^\theta_j=y)M_{xy \rightarrow x'y'} \\
  & = \frac{1}{2}
  `.
\end{align*}
Thus, assuming the agent has model $\theta$ built-in, then Eq.
(\ref{eq:ExampleWork}) determines that the work production for mapping $A\times
y$ to output $A\times y'$ for a particular symbol $y$ is:
\begin{align*}
\left\langle W_{|A \times y, A\times y'} \right\rangle
  = \kB T\left(\ln 2+ \ln \theta^{(y)}_{A \rightarrow A} \right)
  ~.
\end{align*}
Since $A$ is the only memory state and work does not depend on the output
symbol $y'$, the average work produced from an input $y$ is:
\begin{align}
\left\langle W_{|y}\rangle = \langle W_{|A \times y, A\times y'} \right\rangle
  ~.
\end{align}

With the work production expressed for a single input $y_j$, we can now
consider how much work our designed agent harvests from the training data
$y_{0:L}$. Summing the work production of each input yields a simple
expression in terms of the model $\theta$:
\begin{align*}
\left\langle W_{|y_{0:L}} \right\rangle &=\sum_{j=0}^{L-1}
\left\langle W_{|y_j} \right\rangle \\
  & =\sum_{j=0}^{L-1} \kB T
  \left(\ln 2+ \ln \theta^{(y_j)}_{A \rightarrow A} \right) \\
  & = \kB T \left(L\ln 2
  + \ln \prod_{j=0}^{L-1}\theta^{(y_j)}_{A \rightarrow A} \right)
  ~.
\end{align*}
Due to the single causal state, the product within the logarithm simplifies to
the probability of the word given the model $\prod_{j=0}^{L-1}\theta^{(y_j)}_{A
\rightarrow A}=\Pr(Y_{0:L}^\theta=y_{0:L})$. So, the resulting work production
depends on the familiar log-likelihood:
\begin{align*}
\langle W_{|y_{0:L}} \rangle
  &= \kB T \left(L\ln 2+\ell (\theta|y_{0:L}) \right) \\
  &= \langle W^\theta_{|y_{0:L}} \rangle
  ~,
\end{align*}
again, achieving efficient work production, as expected.

\subsection{Maximizing Work for Memoryless Models}

Leveraging the explicit construction for efficient information ratchets, we can
search for the agent that maximizes work from the input string $y_{0:L}$. To
infer a model through work maximization, we label the frequency of $\uparrow$
states in this sequence with $f(\uparrow)$ and the frequency of $\downarrow$
with $f(\downarrow)$. The corresponding log-likelihood of the model is:
\begin{align*}
\ell(\theta|y_{0:L})
  & = \ln \left(\theta^{(\uparrow)}_{A \rightarrow A}\right)^{Lf(\uparrow)}
  \left(\theta^{(\downarrow)}_{A \rightarrow A} \right)^{Lf(\downarrow)} \\
  & = Lf(\uparrow) \ln \left(\theta^{(\uparrow)}_{A \rightarrow A}\right)
  + Lf(\downarrow)\ln \left(\theta^{(\downarrow)}_{A \rightarrow A}\right)
  ~.
\end{align*}
Thus, for the corresponding agent, the work production is:
\begin{align*}
\left\langle W^\theta_{|y_{0:L}} \right\rangle
  & = \kB T \ell(\theta|y_{0:L})+ \kB T L \ln 2 \\
  & = \kB T L  \left(\ln 2
  \! + \! f(\uparrow) \ln \theta^{(\uparrow)}_{A \rightarrow A}
  \! + \! f(\downarrow)\ln\theta^{(\downarrow)}_{A \rightarrow A}\right)
  .
\end{align*}

Selecting from all possible memoryless agents, the model parameters $\theta$
maximizing work production are given by the frequency of symbols in the input:
$f(\uparrow)=  \theta^{(\uparrow)}_{A \rightarrow A}$ and $f(\downarrow) =
\theta^{(\downarrow)}_{A \rightarrow A}$. The resulting work production is:
\begin{align*}
\left\langle W^\theta_{|y_{0:L}} \right\rangle
  = \kB T L  (\ln 2-H[f(\uparrow)])
  ~,
\end{align*}
where $H[f(\uparrow)]$ is the Shannon entropy of binary variable $Y$ with $\Pr(Y=\uparrow)=f(\uparrow)$ measured in nats.

This simple example of learning statistical bias serves to explicitly lay out
the stages of thermodynamic machine learning. The class of models is too
simple, though, to illustrate the full power of the new learning method. That
said, it does confirm that thermodynamic work maximization leads to useful
models of data in the simplest case. As one would expect, the simple agent
found by thermodynamic machine learning discovers the frequency of zeros in the
input and, thus, it learns about its environment. The corresponding work
production is the same as energetic gain of randomizing $L$ bits distributed
according to the frequency $f(\uparrow)$.

However, this neglects the substantial thermodynamic benefits possible with
temporally-correlated environments \cite{Boyd16c}. To illustrate how to extract
this additional energy, a sequel designs and analyzes memoryful agents.


\begin{thebibliography}{}

\bibitem{Cuvi18a}
(Baron)~G. Cuvier.
\newblock {\em Essay on the Theory of the Earth}.
\newblock Kirk and Mercein, New York, 1818.

\bibitem{Berg11a}
H.~Bergson.
\newblock {\em Creative Evolution}.
\newblock Henry Holt and Company, New York, New York, 1907.

\bibitem{Thom61}
D.~W. Thompson.
\newblock {\em On Growth and Form}.
\newblock Cambridge University Press, Cambridge, 1917.

\bibitem{Wien48}
N.~Wiener.
\newblock {\em Cybernetics: Or Control and Communication in the Animal and the
  Machine}.
\newblock MIT, Cambridge, 1948.

\bibitem{Wien88a}
N.~Wiener.
\newblock {\em The Human Use Of Human Beings: {Cybernetics} And Society}.
\newblock Da Capo Press, Cambridge, 1988.

\bibitem{Denn91a}
D.~C. Dennett.
\newblock {\em Consciousness Explained}.
\newblock Little, Brown and Co., New York, New York, 1991.

\bibitem{Goul79a}
S.~J. Gould and R.~Lewontin.
\newblock The spandrels of san marco and the panglossian paradigm: A critique
  of the adaptationist programme.
\newblock {\em Proc. Roy. Soc. Lond. B}, 205(1161):581--598, 1979.

\bibitem{Denn95a}
D.~C. Dennett.
\newblock {\em Darwin's Dangerous Idea: Evolution and the Meanings of Life}.
\newblock Simon and Schuster, New York, New York, 1995.

\bibitem{Mayn98a}
J.~{Maynard-Smith} and E.~Szathmary.
\newblock {\em The Major Transitions in Evolution}.
\newblock Oxford University Press, Oxford, reprint edition, 1998.

\bibitem{Wagn14a}
G.~P. Wagner.
\newblock {\em Homology, Genes, and Evolutionary Innovation}.
\newblock Princeton University Press, Princeton, New Jersey, 2014.

\bibitem{Thom74a}
W.~Thomson.
\newblock Kinetic theory of the dissipation of energy.
\newblock {\em Nature}, pages 441--444, 9 April 1874.

\bibitem{Maxw88a}
J.~C. Maxwell.
\newblock {\em Theory of Heat}.
\newblock Longmans, Green and Co., London, United Kingdom, ninth edition, 1888.

\bibitem{Saga12a}
T.~Sagawa.
\newblock Thermodynamics of information processing in small systems.
\newblock {\em Prog. Theo. Phys.}, 127(1):1--56, 2012.

\bibitem{Parr15a}
J.~M.~R. Parrondo, J.~M. Horowitz, and T.~Sagawa.
\newblock Physics of information.
\newblock {\em Nature Physics}, 11(2):131, 2015.

\bibitem{Shai14a}
S.~Shalev-Shwatrz and S.~Ben-David.
\newblock {\em Understanding Machine Learning: From Theory to Algorithms}.
\newblock Cambridge University Press, 2014.

\bibitem{Hast16a}
T.~Hastie, R.~Tibshirani, and J.~Friedman.
\newblock {\em The Elements of Statistical Learning: Data Mining, Inference,
  and Prediction}.
\newblock Springer Series in Statistics. Springer, New York, New York, second
  edition, 1818.

\bibitem{Meht19a}
P.~Mehta, M.~Bukov, C.-H. Wang, A.~G.~R. Day, C.~Richardson, C.~K. Fisher, and
  D.~J. Schwab.
\newblock A high-bias, low-variance introduction to machine learning for
  physicists.
\newblock {\em Physics Reports}, 810(1-124), 2019.

\bibitem{Lin17a}
H.~W. Lin, M.~Tegmark, and D.~Rolnick.
\newblock Why does deep and cheap learning work so well?
\newblock {\em J. Stat. Phys.}, 168(6):1223--1247, 2017.

\bibitem{Land61a}
R.~Landauer.
\newblock Irreversibility and heat generation in the computing process.
\newblock {\em IBM J. Res. Develop.}, 5(3):183--191, 1961.

\bibitem{Benn87a}
C.~H. Bennett.
\newblock Demons, engines and the {Second Law}.
\newblock {\em Sci. Am.}, 257(5):108--116, 1987.

\bibitem{Szil29a}
L.~Szilard.
\newblock On the decrease of entropy in a thermodynamic system by the
  intervention of intelligent beings.
\newblock {\em Z. Phys.}, 53:840--856, 1929.

\bibitem{Watk93a}
T.~L.~H. Watkins, A.~Rau, and M.~Biehl.
\newblock The statistical mechanics of learning a rule.
\newblock {\em Rev. Mod. Phys.}, 65:499 -- 556, 1993.

\bibitem{Enge01a}
A.~Engel and C.~Van den Broeck.
\newblock {\em Statistical Mechanics of Learning}.
\newblock Cambridge University Press, 2001.

\bibitem{Bell11a}
A.~Bell.
\newblock Learning out of equilibrium.
\newblock In J.~P. Crutchfield and J.~Machta, editors, {\em Santa Fe Institute
  Workshop on Randomness, Structure, and Causality, 9-13 January 2011},
  volume~21, 2011.

\bibitem{Dick15a}
J.~Sohl-Dickstein, E.~A. Weiss, N.~Maheswaranthan, and S.~Ganguli.
\newblock Deep unsupervised learning using nonequilibrium thermodynamics.
\newblock {\em arXiv:1503.03585}, 2015.

\bibitem{Gold17a}
S.~Goldt and U.~Seifert.
\newblock Stocastic thermodynamics of learning.
\newblock {\em Phys. Rev. Lett.}, 118(010601), 2017.

\bibitem{Alem18a}
A.~A. Alemi and I.~Fischer.
\newblock {TherML}: Thermodynamics of machine learning.
\newblock {\em arXiv:1807.04162}, 2018.

\bibitem{Bahr20a}
Y.~Bahri, J.~Kadmon, J.~Pennington, S.~S. Schoenholz, J.~Sohl-Dickstein, and
  S.~Ganguli.
\newblock Statistical mechanics of deep learning.
\newblock {\em Ann. Rev. Cond. Matter Physics}, 11:501--528, 2020.

\bibitem{Boyd16d}
A.~B. Boyd, D.~Mandal, and J.~P. Crutchfield.
\newblock Leveraging environmental correlations: The thermodynamics of
  requisite variety.
\newblock {\em J. Stat. Phys.}, 167(6):1555--1585, 2016.

\bibitem{Mand012a}
D.~Mandal and C.~Jarzynski.
\newblock Work and information processing in a solvable model of {Maxwell's}
  demon.
\newblock {\em Proc. Natl. Acad. Sci. USA}, 109(29):11641--11645, 2012.

\bibitem{Boyd15a}
A.~B. Boyd, D.~Mandal, and J.~P. Crutchfield.
\newblock Identifying functional thermodynamics in autonomous {Maxwellian}
  ratchets.
\newblock {\em New J. Physics}, 18:023049, 2016.

\bibitem{Gold19a}
J.~M. Gold and J.~L. England.
\newblock Self-organized novelty detection in driven spin glasses.
\newblock {\em arXiv:1911.07216}, 2019.

\bibitem{Zhon20a}
W.~Zhong, J.~M. Gold, S.~Marzen, J.~L. England, and N.~Y. Halpern.
\newblock Learning about learning by many-body systems.
\newblock {\em arXiv:2004.03604 [cond-mat.stat-mech]}, 2020.

\bibitem{Reze14a}
D.~Jimenez Rezende, S.~Mohamed, and D.~Wierstra.
\newblock Stochastic backpropagation and approximate inference in deep
  generative models.
\newblock {\em arXiv preprint arXiv:1401.4082}, 2014.

\bibitem{Demp77a}
A.~P. Dempster, N.~M. Laird, and D.~B. Rubin.
\newblock Maximum likelihood from incomplete data via the {EM} algorithm.
\newblock {\em J. Roy. Stat. Soc. Series B}, 39(1):1--22, 1977.

\bibitem{Crut12a}
J.~P. Crutchfield.
\newblock Between order and chaos.
\newblock {\em Nature Physics}, 8(January):17--24, 2012.

\bibitem{Deff2013}
S.~Deffner and C.~Jarzynski.
\newblock Information processing and the second law of thermodynamics: An
  inclusive, {Hamiltonian} approach.
\newblock {\em Phys. Rev. X}, 3:041003, 2013.

\bibitem{Boyd16c}
A.~B. Boyd, D.~Mandal, and J.~P. Crutchfield.
\newblock Correlation-powered information engines and the thermodynamics of
  self-correction.
\newblock {\em Phys. Rev. E}, 95(1):012152, 2017.

\bibitem{Boyd16e}
A.~B. Boyd, D.~Mandal, P.~M. Riechers, and J.~P. Crutchfield.
\newblock Transient dissipation and structural costs of physical information
  transduction.
\newblock {\em Phys. Rev. Lett.}, 118:220602, 2017.

\bibitem{Merh15a}
N.~Merhav.
\newblock Sequence complexity and work extraction.
\newblock {\em J. Stat. Mech.}, page P06037, 2015.

\bibitem{Merh17a}
N.~Merhav.
\newblock Relations between work and entropy production for general
  information-driven, finite-state engines.
\newblock {\em J. Stat. Mech.: Th. Expt.}, 2017:1--20, 2017.

\bibitem{Garn15}
A.~J.~P. Garner, J.~Thompson, V.~Vedral, and M.~Gu.
\newblock Thermodynamics of complexity and pattern manipulation.
\newblock {\em Phys. Rev. E}, 95(042140), 2017.

\bibitem{Deff12a}
S.~Deffer and E.~Lutz.
\newblock Information free energy for nonequilibrium states.
\newblock {\em arXiv:1201.3888}, 2012.

\bibitem{Espo11}
M.~Esposito and C.~van~den Broeck.
\newblock Second law and {Landauer} principle far from equilibrium.
\newblock {\em Europhys. Lett}, 95:40004, 2011.

\bibitem{Croo99a}
G.~E. Crooks.
\newblock Entropy production fluctuation theorem and the nonequilibrium work
  relation for free energy differences.
\newblock {\em Phys. Rev. E}, 60(3), 1999.

\bibitem{Jarz00a}
C.~Jarzynski.
\newblock Hamiltonian derivation of a detailed fluctuation theorem.
\newblock {\em J. Stat. Phys.}, 98(77-102), 2000.

\bibitem{Spec04a}
T.~Speck and U.~Seifert.
\newblock Distribution of work in isothermal nonequilibrium processes.
\newblock {\em Phys. Rev. E}, 70(066112), 2004.

\bibitem{Ray20b}
K.~J. Ray, G.~W. Wimsatt, A.~B. Boyd, and J.~P. Crutchfield.
\newblock Non-markovian momentum computing: Universal and efficient.
\newblock 2020.
\newblock arxiv:2010.01152.

\bibitem{Kolc17a}
A.~Kolchinsky and D.~H. Wolpert.
\newblock Dependence of dissipation on the initial distribution over states.
\newblock {\em J. Stat. Mech.: Th. Expt.}, page 083202, 2017.

\bibitem{Riec20a}
P.~M. Riechers and M.~Gu.
\newblock Initial-state dependence of thermodynamic dissipation for any quantum
  process.
\newblock {\em arXiv:2002.11425}, 2020.

\bibitem{Broo89a}
J.~G. Brookshear.
\newblock {\em Theory of computation: {Formal} languages, automata, and
  complexity}.
\newblock Benjamin/Cummings, Redwood City, California, 1989.

\bibitem{Barn13a}
N.~Barnett and J.~P. Crutchfield.
\newblock Computational mechanics of input-output processes: {Structured}
  transformations and the $\epsilon$-transducer.
\newblock {\em J. Stat. Phys.}, 161(2):404--451, 2015.

\bibitem{Boyd17a}
A.~B. Boyd, D.~Mandal, and J.~P. Crutchfield.
\newblock Thermodynamics of modularity: Structural costs beyond the landauer
  bound.
\newblock {\em Phys. Rev. X}, 8(031036), 2018.

\bibitem{Kirc48a}
G.~R. Kirchhoff.
\newblock {\em Ann. Phys.}, 75:1891, 1848.

\bibitem{Gibb06a}
J.~W. Gibbs.
\newblock {\em The Scientific Papers of J. Willard Gibbs}.
\newblock Longmans, Green, New York, New York, 1906.

\bibitem{Maxw54a}
J.~C. Maxwell.
\newblock {\em A Treatise on Electricity and Magnetism, vol. {I} and {II}}.
\newblock Dover Publications, Inc., New York, New York, third edition, 1954.

\bibitem{Onsa31a}
L.~Onsager.
\newblock Reciprocal relations in irreversible processes, {I}.
\newblock {\em Phys. Rev.}, 37(4):405--426, 1931.

\bibitem{Prig45a}
I.~Prigogine.
\newblock Mod\'eration et transformations irr\'eversibles des syst\`emes
  ouverts.
\newblock {\em Bulletin de la Classe des Sciences, Acad\'emie Royale de
  Belgique}, 31:600--606, 1945.

\bibitem{Prig71b}
I.~Prigogine and P.~Glansdorff.
\newblock {\em Thermodynamic Theory of Structure, Stability and Fluctuations}.
\newblock Wiley-Interscience, London, 1971.

\bibitem{Fala18a}
G.~Falasco, R.~Rao, and M.~Esposito.
\newblock Information thermodynamics of {Turing} patterns.
\newblock {\em Phys. Rev. Let.}, 121:108301, 2018.

\bibitem{Turi52}
A.~M. Turing.
\newblock The chemical basis of morphogenesis.
\newblock {\em Trans. Roy. Soc., Series B}, 237:5, 1952.

\bibitem{Hoyl06a}
R.~Hoyle.
\newblock {\em Pattern Formation: {An} Introduction to Methods}.
\newblock Cambridge University Press, New York, 2006.

\bibitem{Cros09a}
M.~Cross and H.~Greenside.
\newblock {\em Pattern Formation and Dynamics in Nonequilibrium Systems}.
\newblock Cambridge University Press, Cambridge, United Kingdom, 2009.

\bibitem{Heis67a}
W.~Heisenberg.
\newblock Nonlinear problems in physics.
\newblock {\em Physics Today}, 20:23--33, 1967.

\bibitem{Ruel71a}
D.~Ruelle and F.~Takens.
\newblock On the nature of turbulence.
\newblock {\em Comm. Math. Phys.}, 20:167--192, 1971.

\bibitem{Bran83}
A.~Brandstater, J.~Swift, Harry~L. Swinney, A.~Wolf, J.~D. Farmer, E.~Jen, and
  J.~P. Crutchfield.
\newblock Low-dimensional chaos in a hydrodynamic system.
\newblock {\em Phys. Rev. Lett.}, 51:1442, 1983.

\bibitem{Shan48a}
C.~E. Shannon.
\newblock A mathematical theory of communication.
\newblock {\em Bell Sys. Tech. J.}, 27:379--423, 623--656, 1948.

\bibitem{Cove06a}
T.~M. Cover and J.~A. Thomas.
\newblock {\em Elements of Information Theory}.
\newblock Wiley-Interscience, New York, second edition, 2006.

\bibitem{Jame11a}
R.~G. James, C.~J. Ellison, and J.~P. Crutchfield.
\newblock Anatomy of a bit: {Information} in a time series observation.
\newblock {\em CHAOS}, 21(3):037109, 2011.

\bibitem{Shan56b}
C.~E. Shannon.
\newblock The bandwagon.
\newblock {\em IEEE Trans. Info. Th.}, 2(3):3, 1956.

\bibitem{Turi37a}
A.~Turing.
\newblock On computable numbers, with an application to the
  {Entschiedungsproblem}.
\newblock {\em Proc. Lond. Math. Soc.}, 42, 43:230--265, 544--546, 1937.

\bibitem{Ashb57a}
W.~Ross Ashby.
\newblock {\em An Introduction to Cybernetics}.
\newblock John Wiley and Sons, New York, second edition, 1960.

\bibitem{Jayn57a}
E.~T. Jaynes.
\newblock Information theory and statistical mechanics.
\newblock {\em Phys. Rev.}, 106(4):620--630, 1957.

\bibitem{Jayn80a}
E.~T. Jaynes.
\newblock The minimum entropy production principle.
\newblock {\em Annu. Rev. Phys. Chem.}, 31:579--601, 1980.

\bibitem{Kolm58}
A.~N. Kolmogorov.
\newblock A new metric invariant of transient dynamical systems and
  automorphisms in {Lebesgue} spaces.
\newblock {\em Dokl. Akad. Nauk. SSSR}, 119:861, 1958.
\newblock (Russian) Math. Rev. vol. 21, no. 2035a.

\bibitem{Pack80}
N.~H. Packard, J.~P. Crutchfield, J.~D. Farmer, and R.~S. Shaw.
\newblock Geometry from a time series.
\newblock {\em Phys. Rev. Let.}, 45:712, 1980.

\bibitem{Take81}
F.~Takens.
\newblock Detecting strange attractors in fluid turbulence.
\newblock In D.~A. Rand and L.~S. Young, editors, {\em Symposium on Dynamical
  Systems and Turbulence}, volume 898, page 366, Berlin, 1981. Springer-Verlag.

\bibitem{Crut87a}
J.~P. Crutchfield and B.~S. McNamara.
\newblock Equations of motion from a data series.
\newblock {\em Complex Systems}, 1:417 -- 452, 1987.

\bibitem{Bril62a}
L.~Brillouin.
\newblock {\em Science and Information Theory}.
\newblock Academic Press, New York, second edition, 1962.

\bibitem{Benn82}
C.~H. Bennett.
\newblock Thermodynamics of computation---a review.
\newblock {\em Intl. J. Theo. Phys.}, 21:905, 1982.

\bibitem{Saga13}
T.~Sagawa and M.~Ueda.
\newblock Information thermodynamics: Maxwell's demon in nonequilibrium
  dynamics.
\newblock {\em Nonequilibrium Stat. Phys. Small Syst. Fluct. Relations Beyond},
  pages 181--211, nov 2013.

\bibitem{Crut04a}
J.~P. Crutchfield and O.~Gornerup.
\newblock Objects that make objects: {The} population dynamics of structural
  complexity.
\newblock {\em J. Roy. Soc. Interface}, 3:345--349, 2006.

\bibitem{Engl13a}
J.~England.
\newblock Statistical physics of self-replication.
\newblock {\em J. Chem. Phys.}, 139(121923), 2013.

\bibitem{Serr07a}
V.~Serreli, C.-F. Lee, E.~R. Kay, and D.~A. Leigh.
\newblock A molecular information ratchet.
\newblock {\em Nature}, 445(7127):523--537, 2007.

\bibitem{Thom17a}
J.~Thompson, A.~J.~P. Garner, V.~Vedral, and M.~Gu.
\newblock Using quantum theory to simplify input--output processes.
\newblock {\em njp Quant. Info.}, 3(6), 2017.

\bibitem{Loom19a}
S.~Loomis and J.~P. Crutchfield.
\newblock Thermal efficiency of quantum memory compression.
\newblock {\em Phys. Rev. Lett.}, 125:020601, 2020.

\bibitem{Wood18a}
M.~P. Woods, R.~Silva, G.~Putz, S.~Stupar, and R.~Renner.
\newblock Quantum clocks are more accurate than classical ones.
\newblock {\em arXiv preprint arXiv:1806.00491}, 2018.

\bibitem{Daya95a}
P.~Dayan, G.~E. Hinton, R.~M. Neal, and R.~S. Zemel.
\newblock The {Helmholtz} machine.
\newblock {\em Neural computation}, 7(5):889--904, 1995.

\bibitem{LeCu15a}
Y.~LeCun, Y.~Bengio, and G.~Hinton.
\newblock Deep learning.
\newblock {\em Nature}, 521:436--444, May 2015.

\bibitem{Land91a}
R.~Landauer.
\newblock Information is physical.
\newblock {\em Physics Today}, pages 23--29, May 1991.

\bibitem{Crut01a}
J.~P. Crutchfield and D.~P. Feldman.
\newblock Regularities unseen, randomness observed: Levels of entropy
  convergence.
\newblock {\em CHAOS}, 13(1):25--54, 2003.

\bibitem{Stre13a}
C.~C. Strelioff and J.~P. Crutchfield.
\newblock Bayesian structural inference for hidden processes.
\newblock {\em Phys. Rev. E}, 89:042119, 2014.

\bibitem{Wu19a}
T.~Wu and M.~Tegmark.
\newblock Toward an artificial intelligence physicist for unsupervised
  learning.
\newblock {\em Phys. Rev. E}, 100(3):033311, 2019.

\bibitem{Crut92c}
J.~P. Crutchfield.
\newblock The calculi of emergence: Computation, dynamics, and induction.
\newblock {\em Physica D}, 75:11--54, 1994.

\bibitem{Jarz11a}
C.~Jarzynski.
\newblock Equalities and inequalities: Irreversibility and the second law of
  thermodynamics at the nanoscale.
\newblock {\em Annu. Rev. Cond. Matt. Phys.}, 2(1), 2011.

\bibitem{Owen19a}
J.~A. Owen, A.~Kolchinsky, and D.~H. Wolpert.
\newblock Number of hidden states needed to physically implement a given
  conditional distribution.
\newblock {\em New J. Phys.}, 21(013022), 2019.

\end{thebibliography}
\end{document}